THE TWO SAMPLE PROBLEM:

EXACT DISTRIBUTIONS, NUMERICAL SOLUTIONS, SIMULATIONS

by D. E. Chambers, Ph.D.

*Keywords*: Two sample problem, Welch-Aspin solution, Fisher-Behrens problem, nuisance parameter, similarity, the Linnik phenomenon.

*Notation: To avoid ambiguities in the meaning of algebraic expressions, we shall write    a/b. + c/d  to mean  (ad + cb) / bd.  The full-stop or period in bold in the first expression terminates the action of the preceding division sign in the remainder of the expression.*

## § 1.  Introduction.

Let  $X_i$ ,  $i = 1, ... , n$ , be  $n$  independent observations on  random variable  $X$  with  $\mathbf{E}(X) = \mu$  and  $\mathbf{Var}(X) = \sigma^2$ ; such a collection of observations is called a *random sample*.  We define the *mean* of a random sample to be  $\tilde{X} = \Sigma_i X_i / n$ , where  $\mathbf{E}(\tilde{X}) = \mu$  and  $\mathbf{Var}(\tilde{X}) = \sigma^2/n$.   The *sample variance* of the  $n$  observations is defined by  $S^2 = \Sigma_i(X_i - \tilde{X})^2/(n-1)$   where  $\mathbf{E}(S^2) = \sigma^2$.  It is well known that, when  $X$  has the normal distribution  $\mathbf{N}(\mu, \sigma^2)$, the random variables  $\tilde{X}$  and  $S^2$  are independent and are jointly sufficient for  $\mu$  and  $\sigma^2$ .  Also  $\tilde{X}$  has the distribution  $\mathbf{N}(\mu, \sigma^2/n)$  and  $(n-1)S^2/\sigma^2 = U$  has the  $\chi^2$  distribution with  $(n - 1)$  degrees of freedom.

In the two sample problem a test is required to decide whether observations made on each of two normally distributed random variables have different expected values, unaffected by possibly different variability in the two sets of observations.  More specifically, let a random sample of $n_1$ independent observations  $X_{1i}$ ,  $i = 1 , ... , n_1$ ,  $\mathbf{E}(X_{1i}) = \mu_1$ , $\mathbf{Var}(X_{1i}) = \sigma_1^2$,  be made on the one of the normal random variables and a sample of $n_2$ independent observations  $X_{2j}$ ,  $j = 1 , ... , n_2$ ,  $\mathbf{E}(X_{2j}) = \mu_2$ ,  $\mathbf{Var}(X_{2j}) = \sigma_2^2$ , be made on the other normal random variable: we require a statistical test of the null hypothesis: H$_o$: $\mu_1 = \mu_2$,  versus  H$_1$: $\mu_1 \neq \mu_2$  that is unaffected by the unknown value of the 'nuisance parameter'  $\sigma_1^2/\sigma_2^2 = \zeta$ .

To make a statistical decision an appropriate statistic is required.  If the random variables  $X_1$  and  $X_2$  have normal distributions and the values of  $\sigma_1^2$  and  $\sigma_2^2$  were known, then a test of H$_o$ versus H$_1$ could be based on the random variable

$$(\tilde{X}_1 - \tilde{X}_2)/(\sigma_1^2/n_1 \,\mathbf{.} + \sigma_2^2/n_2)^{1/2} \sim \mathbf{N}(0,1)$$

Since  $\sigma_1^2$  and  $\sigma_2^2$  are assumed to be unknown, the factor $(\sigma_1^2/n_1 \,\mathbf{.} + \sigma_2^2/n_2)^{1/2}$ in this random variable must be replaced by a statistical estimator.  Let  $\tilde{X}_1$ and  $\tilde{X}_2$  be the sample mean of the first and second random samples respectively, and let $S_1^2$ and  $S_2^2$  be their respective sample variances.  Since

$$\mathbf{Var}(\tilde{X}_1 - \tilde{X}_2) = \mathbf{E}(S_1^2/n_1 \,\mathbf{.} + S_2^2/n_2) = \sigma_1^2/n_1 \,\mathbf{.} + \sigma_2^2/n_2 ,$$

it follows that  $S_1^2/n_1 \,\mathbf{.} + S_2^2/n_2$  is a statistical estimator of  $\sigma_1^2/n_1 \,\mathbf{.} + \sigma_2^2/n_2$, and hence the value of the standardized  random variable $(\tilde{X}_1 - \tilde{X}_2)/(\sigma_1^2/n_1 \,\mathbf{.} + \sigma_2^2/n_2)^{1/2}$ is estimated by the statistic $V = (\tilde{X}_1 - \tilde{X}_2)/(S_1^2/n_1 \,\mathbf{.} + S_2^2/n_2)^{1/2}$ .

At the same time account must be taken of the 'nuisance' parameter  $\sigma_1^2/\sigma_2^2 = \zeta$ . Since $\mathbf{E}(S^2) = \sigma^2$  we see that  $S_1^2/S_2^2 = Z$  is a statistical estimator  $\zeta = \sigma_1^2/\sigma_2^2$.

Now consider the independent random variables $U_1 = v_1 S^2_1/\sigma^2$, $U_2 = v_2 S^2_2/\sigma^2$ with $\chi^2$ distributions with $v_1$, $v_2$ degrees of freedom, respectively, where $v_1 = n_1 - 1$ and $v_2 = n_2 - 1$. Therefore

$$Z = S_1^2/S_2^2 = \sigma_1^2 U_1/v_1 \,.\, / \sigma_2^2 U_2/v_2 = v_2\sigma_1^2/v_1\sigma_2^2 \,.\, U_1/U_2 ,$$

and since the probability distribution of the random variable $v_2/v_1 \,.\, U_1/U_2$ is, by definition, the $\mathbf{F}(v_1, v_2)$ distribution, the probability density function of the random variable $Z$ is

$$1/B(\tfrac{1}{2} v_1, \tfrac{1}{2}v_2) \,.\, z^{\frac{1}{2} v_1 - 1} \zeta^{\frac{1}{2}v_2}/(v_2\zeta + v_1 z)^{\frac{1}{2} v} ,$$

where $v = v_1 + v_2$, see Appendix 1.

We shall consider the test due to Welch and the different test due to Fisher and Behrens. Both these tests specify test criteria for the statistic $V$ that are functions $v_\alpha(z)$ of $z$ for a nominated significance level $\alpha$. In the one-tailed test, $H_0 : \mu_1 \leq \mu_2$ versus $H_1: \mu_1 > \mu_2$ is tested at (nominal) significance level $\alpha$ using the test criterion $v_\alpha(z)$ and $H_0$ is rejected at this level if $V > v_\alpha(Z)$. In the two-tailed test of $H_0: \mu_1 = \mu_2$ versus $H_1: \mu_1 \neq \mu_2$ is tested at (nominal) significance level $2\alpha$, $H_0$ is rejected at this level if $|V| > v_\alpha(z)$.

Since the sample variance ratio $z$ is unbounded above it is not suitable variable for tabulating test criteria, however two alternative statistics, each with a finite range of values, have been introduced, namely $c = n_2 z/(n_1 + n_2 z)$ (or $C = n_2 Z/(n_1 + n_2 Z)$), due to Welch, and $\theta = \tan^{-1}(n_2 z/n_1)^{\frac{1}{2}}$ (or $\Theta = \tan^{-1}(n_2 Z/n_1)^{\frac{1}{2}}$), due to Fisher. Notice that $c = \sin^2 \theta$. We define the corresponding population parameters as $\gamma = n_2\zeta/(n_1 + n_2\zeta)$ and $\psi = \tan^{-1}(n_2\zeta/n_1)^{\frac{1}{2}}$.

A test criterion $v_\alpha(z)$ is said to be '*ideal*', or *similar*, if probability of Type-I error when $H_0: \mu_1 = \mu_2$ true is $2\alpha$ for all values of the 'nuisance' parameter $\zeta$. We shall denote an 'ideal' criterion with bold type, i.e. $\boldsymbol{v}_\alpha(z)$ represents an 'ideal' criterion.

Some properties of an 'ideal' criterion $\boldsymbol{v}_\alpha(z)$.

i.  Since $\sigma_1 \to 0$ implies that $S_1 \to 0$, $Z \to 0$ and $\tilde{X}_1 \to \mu$, it follows, in this limit, that $V \to (\mu - \tilde{X}_2)/(S_2^2/n_2)^{\frac{1}{2}} = (v_2)^{\frac{1}{2}} [(n_2)^{\frac{1}{2}}(\mu - \tilde{X}_2) / \sigma_2][ (v_2)^{\frac{1}{2}} S_2 /\sigma_2]$, where $(n_2)^{\frac{1}{2}} (\mu - \tilde{X}_2)/\sigma_2 \sim \mathbf{N}(0,1)$ and $v_2 S^2_2/\sigma_2^2 \sim \chi^2(v_2)$. Therefore the distribution of $V$ tends to the Student-$t$ with $v_2$ degrees of freedom as $\sigma_1 \to 0$ i.e.

$$\boldsymbol{v}_\alpha(z)_{z \to 0} = t_{v_2}(\alpha) , \text{ where } \boldsymbol{S}_{2}(t_{v_2}(\alpha)) = 1 - \alpha .$$

Similarly $\boldsymbol{v}_\alpha(z)_{z \to \infty} = t_{v_1}(\alpha)$, where $\boldsymbol{S}_{v_1}(t_{v_1}(\alpha)) = 1 - \alpha$.

ii  It is shown in § 3 that the random variable

$$T_v = (v)^{\frac{1}{2}} (\tilde{X}_1 - \tilde{X}_2)/(\zeta/n_1 \,.\, + 1/n_2)^{\frac{1}{2}} (v_1 S_1^2/\zeta \,.\, + v_2 S_2^2)^{\frac{1}{2}}, \quad v = v_1 + v_2 ,$$

has the Student-$t$ distribution with $v$ degrees of freedom. Now when $\zeta = n_1 v_1/n_2 v_2$: we see that $V \equiv T_v$ for all $Z$. Therefore, if $\boldsymbol{v}_\alpha(z) > t_v(\alpha)$ for all $z$, $\mathbf{Pr}\{V < \boldsymbol{v}_a(z)\} > \mathbf{Pr}\{T_v < t_v(\alpha)\} = 1 - \alpha$, which is a contradiction since, by definition, $\mathbf{Pr}\{V \leq \boldsymbol{v}_a(z)\} = 1 - \alpha$. Therefore the function $\boldsymbol{v}_\alpha(z) < t_v(\alpha)$ for some $z$ if $\boldsymbol{v}_\alpha(z)$ is to be 'ideal'.

## § 2. The probability distribution of the statistic $V$.

**Lemma 1.** The conditional probability distribution of the random variable

$$n_1 n_2 (S_1^2/n_1 + S_2^2/n_2) \, (v_1 Z/\sigma_1^2 + v_2/\sigma_2^2) \, /(n_2 Z + n_1) \,,$$

given $Z = z$, is the central $\chi^2$ distribution with $v = v_1 + v_2$ degrees of freedom, where $v_1 = n_1 - 1$ and $v_2 = n_2 - 1$.

**Proof.** Assume that the random variables

$$U_1 = v_1 S_1^2/\sigma_1^2 \,, \qquad U_2 = v_2 S_2^2/\sigma_2^2$$

are independent and have $\chi^2$ distributions with respective degrees of freedom $v_1$ and $v_2$. Determination of the joint probability distribution of transformed random variables

$$W = S_1^2/n_1 + S_2^2/n_2 \,, \qquad Z = S_1^2/S_2^2 \,,$$

establishes a proof. $U_1$ and $U_2$ are functions of $W$ and $Z$ given by

$$U_1 = v_1 W/\sigma_1^2 (1/n_1 + 1/Z\,n_2) = W \, h_1(Z) \,,$$

and

$$U_2 = v_2 W/\sigma_2^2 (Z/n_1 + 1/n_2) = W \, h_2(Z) \,,$$

where the Jacobian of this transformation has the functional form

$$J = \partial(u_1, u_2)/\partial(w, z) = w \, g(z).$$

Since the random variables $U_1$ and $U_2$ are independent and have $\chi^2$ distributions with $v_1$ and $v_2$ degrees of freedom, respectively, it follows that joint probability density function of the random variables $W, Z$ has the functional form

$$w^{\frac{1}{2}(v_1 + v_2) - 1} \exp(- \tfrac{1}{2} w \, (h_1(z) + h_2(z))) \, K(z) \,, \quad w > 0, \; z > 0 \,.$$

Therefore the conditional probability density function of the random variable

$$W \, (h_1(Z) + h_2(Z)) = W \, n_1 n_2 \, [v_1 Z/\sigma_1^2 + v_2/\sigma_2^2] \, /(n_2 Z + n_1)$$

must be the probability density function of the $\chi^2$ distribution with $v = v_1 + v_2$ degrees of freedom. (See Appendix 2 for more details.) ●

Using the result of Lemma 1 it is easy to prove

**Lemma 2.** Consider the statistic $V = (\tilde{X}_1 - \tilde{X}_2)/(S_1^2/n_1 + S_2^2/n_2)^{1/2}$. The conditional probability distribution of the random variable

$$V / [n_1 n_2 (v_1 Z/\zeta + v_2)(\zeta/n_1 + 1/n_2)/v(n_2 Z + n_1)]^{1/2} = V/K_{Z,\zeta}$$

given $Z = z$, is the Student-$t$ distribution with $v = v_1 + v_2$ degrees of freedom and non-centrality parameter

$$\delta = (\mu_1 - \mu_2)/(\sigma_1^2/n_1 + \sigma_2^2/n_2)^{1/2}.$$

**Proof.** $[(\tilde{X}_1 - \tilde{X}_2) - (\mu_1 - \mu_2)]/(\sigma_1^2/n_1 + \sigma_2^2/n_2)^{1/2} = \varphi \sim \mathbf{N}(0,1)$. Therefore, when $H_o$ is true, the conditional probability distribution of the random variable

$$v^{1/2}(\tilde{X}_1 - \tilde{X}_2)/(\sigma_1^2/n_1 + \sigma_2^2/n_2)^{1/2} W^{1/2} [n_1 n_2 (v_1 Z/\sigma_1^2 + v_2/\sigma_2^2)/(n_2 Z + n_1)]^{1/2},$$

given $Z$, is the Student-$t$ distribution with $v$ degrees of freedom. But this random variable can be put in the form

$$v^{1/2} V/(\sigma_1^2/n_1 + \sigma_2^2/n_2)^{1/2} [n_1 n_2 (v_1 Z/\sigma_1^2 + v_2/\sigma_2^2)/(n_2 Z + n_1)]^{1/2} =$$

$$V / [n_1 n_2 (v_1 Z/\zeta + v_2)(\zeta/n_1 + 1/n_2)/v(n_2 Z + n_1)]^{1/2} = V/K_{Z,\zeta}.$$

It follows that $\mathbf{Pr}\{V \leq v \mid Z=z, \zeta\} = \mathbf{Pr}\{T_v \leq v/K_{z\,\zeta}\} = S_v(v/K_z\,\zeta)$

$$= 1/B(\tfrac{1}{2}, \tfrac{1}{2}v) \int_{-\infty}^{v/K_{z,\zeta}} dt/v^{1/2}(1 - t^2/v)^{\frac{1}{2}(v+1)},$$

where $K_{z,\zeta}^2 = (v_1 z/\zeta + v_2)(n_2\zeta + n_1)/v(n_2 z + n_1)$.       ●

Since the probability distribution of $Z$ is a scaled version of the $\mathbf{F}(v_1, v_2)$ distribution (see § 1 and Appendix 1), a simple conditional probability argument leads to the main theoretical result of this article.

**Theorem.** Under the usual assumptions the probability of the event $\{V \leq v(Z) \mid \zeta\}$ is given by

$$v_1^{\frac{1}{2}v_1} v_2^{\frac{1}{2}v_2} / B(\tfrac{1}{2}v_1, \tfrac{1}{2}v_2) \int_0^\infty S_v(v(z)/K_z\,\zeta)\ \zeta^{\frac{1}{2}v_2} z^{\frac{1}{2}v_1 - 1}/(v_2\zeta + v_1 z)^{\frac{1}{2}v}\ dz$$

where $S_v(\cdot)$ is the cumulative probability distribution function of the Student-$t$ distribution with $v$ degrees of freedom, and $B(\tfrac{1}{2}v_1, \tfrac{1}{2}v_2)$ is the $\beta$ function. ●

**Corollary.** An alternative to the expression above that involves integration over a finite interval is

$$\mathbf{Pr}\{V \leq v_c(C) \mid \gamma\} = 1/B(\tfrac{1}{2}v_1, \tfrac{1}{2}v_2) \int_0^1 S_v(v_c(v\,\gamma\ x\ K_{x,\gamma}^2/v_1)/K_{x,\gamma})\ x^{\frac{1}{2}v_1 - 1}(1 - x)^{\frac{1}{2}v_2 - 1}dx,$$

where $K_{x,\gamma}^2 = v_1 v_2/v [v_1(1-\gamma)(1-x) + v_2\gamma x]$. Compared with the previous integral expression the statistic $z$ is replaced by $c = n_2 z/(n_1 + n_2 z)$ and the variance ratio $\zeta$ by its alternative $\gamma = n_2\zeta/(n_1 + n_2\zeta)$.

**Proof.** Replacing the variable of integration by means of the substitution

$$z = \zeta \, v_2 \, x \, / v_1 (1 - x)$$

in the first integral expression transforms the range of integration from $(0, \infty)$ to $(0, 1)$. It only remains to show that the argument of the function $S_v(\cdot)$ is as stated. Firstly

$$K_{z,\zeta}^2 = (v_1 z / \zeta + v_2)(\zeta n_2 + n_1) / v \, (n_2 z + n_1), \quad \text{where} \quad z = \zeta v_2 x / v_1(1 - x),$$

$$= v_1 \, v_2 \, (n_2 \zeta + n_1) / v \, [n_2 \, v_2 \, \zeta x + n_1 v_1(1 - x)], \quad \text{and since} \quad \zeta = n_1 \, \gamma \, / \, n_2(1 - \gamma),$$

$$= v_1 \, v_2 \, / v \, (v_2 \, x \, \gamma + v_1(1 - x)(1 - \gamma)) = K_{x,\gamma}^2.$$

The relationship between different versions of the same test criterion are given by the equations

$$v_c(c) = v_z(z) = v_z(n_1 \, c \, / \, n_2(1 - c)) \quad \text{or} \quad v_z(z) = v_c(c) = v_c(n_2 z / (n_1 + n_2 z))$$

where $z = n_1 \, c \, / \, n_2(1 - c)$ or, inversely, $c = n_2 \, z \, / (n_1 + n_2 z)$. By substituting the new variable of integration, $x$, into $v_c(n_2 z / (n_1 + n_2 z))$ and then replacing $\zeta$ with $\gamma$ using $\zeta = n_1 \, \gamma \, / \, n_2(1 - \gamma)$, we obtain

$$v_c(n_2 z / (n_1 + n_2 z)) = v_c(n_2 \, \zeta \, v_2 \, x \, / v_1(1 - x)[n_1 + n_2 \, \zeta \, v_2 \, x \, / v_1(1 - x)]$$

$$= v_c(n_2 \, n_1 \, \gamma \, / \, n_2(1 - \gamma) \cdot v_2 \, x \, / \, [v_1(1 - x)n_1 + n_2 \, n_1 \, \gamma \, / \, n_2(1 - \gamma) \, v_2 \, x]$$

$$= v_c( \, v_2 \, \gamma \, x \, / \, [v_1(1 - x)(1 - \gamma) + \, v_2 \, x \, \gamma \,]) = v_c(v \, \gamma \, x \, K_{x,\gamma}^2 \, / \, v_1) \qquad \bullet$$

(As all the computed test criteria were calculated in terms of the Fisher statistic $\theta$, the formulae of the Theorem and the Corollary were never actually used in computations. Rather, if the Fortran NAG integration sub-routine called for the value of the integrand at $x$, $x$ was converted to the equivalent value of $z$ using $z = \zeta v_2 x / v_1(1 - x)$, and $v(z)$ was found from a table of the test criterion $v(\theta) = v_\theta(\theta)$ which was interpolated for a value at $\theta = \tan^{-1}(n_2 z / n_1)^{1/2}$. The required integrand at $x$ would then be

$$S_v(v(\theta) / K_{z,\zeta}) \, x^{\frac{1}{2} v_1 - 1} (1 - x)^{\frac{1}{2} v_2 - 1}. \, )$$

The limiting form of $\mathbf{Pr}\{V \le v_c(C) | \gamma\}$ as $v_1 \to \infty$ is important in the compilation of tables. One form of this limit is

$$\mathbf{Pr}\{V \le v_c(C) | \gamma\} = (\tfrac{1}{2} v_2)^{\frac{1}{2} v_2} / \Gamma(\tfrac{1}{2} v_2) \cdot \int_o^\infty \Phi(t \, v_c(\gamma / (1 - \gamma)t + \gamma)) \, t^{\frac{1}{2} v_2 - 1} \, e^{-\frac{1}{2} v_2 t} \, dt \, ,$$

where $\Phi$ is the cumulative probability distribution function of the standardized normal distribution.

**Proof.** Setting $z = \zeta / t$ in $\zeta^{\frac{1}{2} v_2} z^{\frac{1}{2} v_1 - 1} / (v_2 \zeta + v_1 \, z)^{\frac{1}{2} v} \cdot dz$ gives

$$\zeta^{\frac{1}{2} v_2} (\zeta / t)^{\frac{1}{2} v_1 - 1} / \zeta^{\frac{1}{2} v} \, (v_2 + v_1 / t)^{\frac{1}{2} v} \cdot -\zeta / t^2 \, dt =$$

$$= [(v_2 \, t / v_1 + 1)^{v_1}]^{-\frac{1}{2}} \, v_1^{-\frac{1}{2} v_1} \, (v_2 \, t / v_1 + 1)^{-\frac{1}{2} v_2} \, v_1^{-\frac{1}{2} v_2} \, t^{\frac{1}{2} v_2 - 1} \, dt$$

$$\to \exp(v_2 t)^{-\frac{1}{2}} \, t^{\frac{1}{2} v_2 - 1} \, v_1^{-\frac{1}{2} v} \quad \text{as} \quad v_1 \to \infty \, .$$

The same substitution into $K_{z,\zeta}{}^2 = (v_1 z/\zeta + v_2)(n_2\zeta + n_1)/v(n_2 z + n_1)$ gives

$$(v_1/t + v_2)(n_2\zeta + n_1)/v(n_2\zeta/t + n_1) \to 1/t \text{ as } v_1 \to \infty.$$

Similarly $v(z) = v_c(n_2 z/(n_1 + n_2 z)) = v_c(n_2\zeta/t(n_1 + n_2\zeta/t)) = v_c(n_2\zeta/(tn_1 + n_2\zeta))$

$$= v_c(n_2 \, n_1 \, \gamma \, / \, n_2(1-\gamma)(tn_1 + n_2 \, n_1 \, \gamma \, / \, n_2(1-\gamma)))$$

$$= v_c(n_2 \, n_1 \, \gamma \, /( \, n_2(1-\gamma)tn_1 + n_2 \, n_1 \, \gamma \, ))$$

$$= v_c( \, \gamma \, /[(1-\gamma)t + \gamma]) \, ,$$

which expression is functionally independent of the sample sizes.

The stated result follows, since $\int_0^{\infty} t^{\frac{1}{2}v_2 - 1} \, e^{-\frac{1}{2}v_2 t} \, dt = \Gamma(\frac{1}{2}v_2) \, /(\frac{1}{2}v_2)^{\frac{1}{2}v_2}$ .

§ 3.  <u>An alternative derivation of the probability distribution of the statistic $V$, and other results.</u>

We shall again derive the most important results concerning the statistic $V$ using a slightly different method, and then apply the same method to obtain theoretical expressions for the performance of other statistics of interest in the two-sample problem.

Introduce the independent random variables $U_1$, $U_2$ with $\chi^2$ distributions with $v_1$ and $v_2$ degrees of freedom, respectively, and set $Y = U_1 + U_2 \sim \chi^2$, and $Z = \sigma_1{}^2 U_1 v_2 / \sigma_2{}^2 U_2 v_1$. It is easy to show that the random variables $Y$ and $Z$ are independent. Solving the equations $Y = U_1 + U_2$, $Z = \sigma_1{}^2 U_1 v_2 / \sigma_2{}^2 U_2 v_1$ for $U_1$ and $U_2$ gives

$$U_1 = Y/(1 + \sigma_1{}^2 v_2 / \sigma_2{}^2 v_1 Z) = \sigma_2{}^2 v_1 \, Y \, Z \, /(\sigma_2{}^2 v_1 Z + \sigma_1{}^2 v_2) = v_1 \, Y \, Z \, /(v_1 Z + v_2 \, \zeta) \, ,$$

$$U_2 = \sigma_1{}^2 v_2 \, Y \, /(\sigma_2{}^2 v_1 Z + \sigma_1{}^2 v_2) = \zeta \, v_2 \, Y \, /(v_1 Z + v_2 \zeta) \, .$$

Now consider $(S_1{}^2/n_1 + S_2{}^2/n_2) = \sigma_1{}^2 \, U_1/v_1 \, n_1 + \sigma_2{}^2 \, U_2/v_2 \, n_2$

$$= \sigma_1{}^2/n_1 \, Y \, Z \, /(v_1 Z + v_2 \, \zeta) + \sigma_2{}^2/n_2 \, \zeta \, Y \, /(v_1 Z + v_2\zeta)$$

$$= Y \, (\sigma_1{}^2/n_1 \, Z + \sigma_2{}^2/n_2 \, \zeta) \, /(v_1 Z + v_2 \, \zeta) \, .$$

Hence the statistic $V = (\tilde{X}_1 - \tilde{X}_2)/(S_1{}^2/n_1 + S_2{}^2/n_2)^{\frac{1}{2}}$ can be put in the form

$$V = (\tilde{X}_1 - \tilde{X}_2)/[ \, Y \, (\sigma_1{}^2/n_1 \, Z + \sigma_2{}^2/n_2 \, \zeta) \, /(v_1 Z + v_2 \, \zeta)]^{\frac{1}{2}}$$

$$= (\tilde{X}_1 - \tilde{X}_2)/Y^{\frac{1}{2}} \, [(\sigma_1{}^2 n_2 \, Z + \sigma_2{}^2 n_1 \, \zeta) \, / \, n_1 n_2 (v_1 Z + v_2 \, \zeta)]^{\frac{1}{2}}$$

$$= v^{\frac{1}{2}}(\tilde{X}_1 - \tilde{X}_2)/Y^{\frac{1}{2}} \, [(v_1 Z + v_2 \, \zeta) \, (n_1 n_2) \, / \, v \, (\sigma_1{}^2 n_2 \, Z + \sigma_2{}^2 n_1 \, \zeta)]^{\frac{1}{2}} \, .$$

But $X_\varphi = (\tilde{X}_1 - \tilde{X}_2) / (\sigma_1^2/n_1. + \sigma_2^2/n_2)^{1/2} \sim \mathbf{N}(0,1)$ and substitution of this into the previous expression gives

$$\nu^{1/2} X_\varphi / Y^{1/2} \cdot [(\sigma_1^2/n_1. + \sigma_2^2/n_2)(n_1 n_2)(\nu_1 Z + \nu_2 \zeta) / \nu(\sigma_1^2 n_2 Z + \sigma_2^2 n_1 \zeta)]^{1/2}$$

$$= T_\nu [(\sigma_1^2 n_2 + \sigma_2^2 n_1)(\nu_1 Z + \nu_2 \zeta) / \nu(\sigma_1^2 n_2 Z + \sigma_2^2 n_1 \zeta)]^{1/2}$$

$$= T_\nu [(\zeta n_2 + n_1)(\nu_1 Z + \nu_2 \zeta) / \nu \zeta(n_2 Z + n_1)]^{1/2} ,$$

i.e. $V = T_\nu K_{Z,\zeta}$, where $K_{Z,\zeta}^2 = (\zeta n_2 + n_1)(\nu_1 Z + \nu_2 \zeta) / \nu \zeta (n_2 Z + n_1)$. Cf. § 2, Lemma 2 and $T_\nu$ has the Student-$t$ distribution with $\nu$ degrees of freedom.

In a similar way we arrive at the following

Generalisation. Replace $(S_1^2/n_1. + S_2^2/n_2)$ in the above analysis with $S_1^2 f_1(n_1,n_2) + S_2^2 f_2(n_1,n_2)$. Then the statistic $(\tilde{X}_1 - \tilde{X}_2)/[S_1^2 f_1(n_1,n_2) + S_2^2 f_2(n_1,n_2)]^{1/2}$ can be put in the form
$$(\tilde{X}_1 - \tilde{X}_2)/[f_1(n_1,n_2)\sigma_1^2/\nu_1^2 X Y + f_2(n_1,n_2)\sigma_2^2/\nu_2 (1 - X) Y]^{1/2} =$$

$$\nu^{1/2} X_\varphi / Y^{1/2} \cdot [(n_1 + n_2\zeta)/n_1 n_2]^{1/2} / [\nu (f_1(n_1,n_2)\zeta X/\nu_1. + f_2(n_1,n_2)(1 - X)/\nu_2)]^{1/2} = T_\nu / k_{X,\zeta} ,$$

where $k_{X,\zeta}^2 = \nu[f_1(n_1,n_2) \zeta X /\nu_1. + f_2(n_1,n_2)(1 - X)/\nu_2] n_1 n_2/(n_1 + n_2\zeta)$, or on eliminating $\zeta$ in favour of $\gamma$ by means of the substitution $\zeta = \nu_1\gamma /\nu_2(1 - \gamma)$, we have the alternative form

$$k_{X,\zeta}^2 = \nu [f_1(n_1,n_2) \gamma X/\nu_1. + f_2(n_1,n_2)(1 - \gamma)(1 - X)/\nu_2] = k_{X,\gamma}^2.$$

Here the random variable $X$ has the $\mathbf{B}(\tfrac{1}{2}\nu_1, \tfrac{1}{2}\nu_2)$ distribution. Since $Y$ and $Z$ are independent, and since $Z = \zeta \nu_2 X/\nu_1(1 - X)$ it follows that $Y$ and $X$ are also independent random variables.

{Check: For the statistic $V$ we have $f_1(n_1,n_2) = 1/n_1$, $f_2(n_1,n_2) = 1/n_2$, hence

$$k_\zeta x^2 = [\zeta X /\nu_1 n_1. + (1 - X)/\nu_2 n_2] \nu n_1 n_2/(n_1 + n_2\zeta)$$

$$= [\zeta Z/(\nu_1 Z + \zeta \nu_2) n_1. + \zeta/(\nu_1 Z + \zeta \nu_2)n_2] \nu n_1 n_2/(n_1 + n_2\zeta)$$

$$= (Z/n_1 + 1/n_2) \zeta \nu n_1 n_2/(\nu_1 Z + \zeta \nu_2)(n_1 + n_2\zeta)$$

$$= (Zn_2 + n_1) \zeta \nu/(\nu_1 Z + \zeta \nu_2)(n_1 + n_2\zeta) = 1/K_{Z,\zeta}^2 ,$$

which agrees with previous derivations.)

By repetition of the steps which lead to the theorem in §2, exact expressions for the performance of all statistics of the type $(\tilde{X}_1 - \tilde{X}_2)/[S_1^2 f_1(n_1,n_2) + S_2^2 f_2(n_1,n_2)]^{1/2}$ are easily obtained. For example, consider the statistic obtained from least squares theory for the comparison of two means:

$$T(\zeta) = (\nu)^{1/2} (\tilde{X}_1 - \tilde{X}_2)/(\zeta /n_1. + 1/n_2)^{1/2} (\nu_1 S_1^2/\zeta. + \nu_2 S_2^2)^{1/2}$$

where, normally, $\boldsymbol{\zeta}$ will be given the value of $\zeta$ if the value of $\zeta$ is known. In the general case

$$f_1(n_1,n_2) = v_1/v. \,(1/n_1. + 1/n_2\,\boldsymbol{\zeta}) \;\;, \;\;\; f_2(n_1,n_2) = v_2/v. \,(\boldsymbol{\zeta}\,/n_1. + 1/n_2) \;,$$

and hence

$$k_{X,\gamma}{}^2 \;=\; (1 + n_1/n_2\boldsymbol{\zeta})\,\gamma X + (1 + n_2\boldsymbol{\zeta}/n_1)(1 - \gamma)(1 - X) \;,$$

implying

$$\mathbf{Pr}\{T(\boldsymbol{\zeta}) \le t_v(\alpha) \mid \gamma\} \;=\; 1/\,\mathrm{B}(\tfrac{1}{2}\,v_1,\tfrac{1}{2}\,v_2) \,.\, \int_o^1 S_v(t_v(\alpha)\,k_{X,\gamma}) \;\; x^{\frac{1}{2}\,v_1 - 1}(1 - x)^{\frac{1}{2}\,v_2 - 1}dx.$$

If $\boldsymbol{\zeta} = \zeta$ then

$$T(\boldsymbol{\zeta}) = (v)^{\frac{1}{2}}\,(\tilde{X}_1 - \tilde{X}_2)/(\,\sigma_1{}^2/n_1. + \sigma_2{}^2/n_2)^{\frac{1}{2}}\,(v_1\,S_1{}^2/\sigma_1{}^2. + v_2\,S_2{}^2/\sigma_2{}^2)^{\frac{1}{2}} \;,$$

and since $\boldsymbol{\zeta} = n_1\gamma/n_2(1 - \gamma)$, we have $(1 + n_1/n_2\,\boldsymbol{\zeta}) = 1 + (1 - \gamma)/\gamma = 1/\gamma$ and $(1 + n_2\,\boldsymbol{\zeta}/n_1) = 1/(1 - \gamma)$ ; consequently $k_{X,\gamma}{}^2 = X + (1 - X) = 1$. Therefore, when $\boldsymbol{\zeta} = \zeta$,

$$\mathbf{Pr}\{T(\boldsymbol{\zeta}) \le t_v(\alpha) \mid \gamma\} \;=\; 1/\,\mathrm{B}(\tfrac{1}{2}\,v_1,\tfrac{1}{2}\,v_2) \,.\, \int_o^1 S_v(t_v(\alpha)) \;\; x^{\frac{1}{2}\,v_1 - 1}(1 - x)^{\frac{1}{2}\,v_2 - 1}dx,$$

$$=\; S_v(t_v(\alpha)) \int_o^1 x^{\frac{1}{2}\,v_1 - 1}(1 - x)^{\frac{1}{2}\,v_2 - 1}dx/\,\mathrm{B}(\tfrac{1}{2}\,v_1,\tfrac{1}{2}\,v_2) \;=\; S_v(t_v(\alpha)) \;=\; 1 - \alpha$$

as required

Power of the statistics $T(\zeta)$ and $V$.

The rejection region of the $2\alpha$ sized two-tailed test of $H_o$: $\mu_1 = \mu_2$ versus $H_1$: $\mu_1 \ne \mu_2$ using the statistic $T(\zeta)$ is given by

$$\mathbf{Pr}\{\,T(\zeta) < -\,t_v(\alpha)\} + \mathbf{Pr}\{\,T(\zeta) > t_v(\alpha)\} \;=\; 1 + \mathbf{Pr}\{\,T(\zeta) < -\,t_v(\alpha)\} \,\textbf{-}\, \mathbf{Pr}\{\,T(\zeta) < t_v(\alpha)\}$$

$$=\; 1 + S_v\{-\,t_v(\alpha)\} - S_v\{t_v(\alpha)\} \;=\; 2\alpha$$

and the power of this test is

$$1 + S_v\{-\,t_v(\alpha),\,\delta\} - S_v\{t_v(\alpha),\,\delta\} \quad (1)$$

where $S_v(t,\delta)$ is the cumulative probability distribution function of the non-central Student-$t$ distribution with $v$ degrees of freedom and non-centrality parameter $\delta = (\mu_1 - \mu_2)\,/(\sigma_1{}^2/n_1. + \sigma_2{}^2/n_2)^{\frac{1}{2}}$, which is identical to that of $V$.

The rejection region of the $2\alpha$ sized two-tailed test of $H_o$: $\mu_1 = \mu_2$ versus $H_1$: $\mu_1 \ne \mu_2$ using the statistic $V$ is given by

$$\mathbf{Pr}\{V < -\,\boldsymbol{v}_\alpha(z)\} + \mathbf{Pr}\{V > \boldsymbol{v}_\alpha(z)\} \;=\; 1 + \mathbf{Pr}\{V < -\,\boldsymbol{v}_\alpha(z)\} \,\textbf{-}\, \mathbf{Pr}\{V < \boldsymbol{v}_\alpha(z)\} \;=\; 2\alpha,$$

and the power function of this test, for a given $\boldsymbol{\zeta}$, is equal to

$$1 + c \int_o^\infty (S_v(-\,\boldsymbol{v}_\alpha(z)/K_z\,_\zeta,\,\delta) - S_v(\boldsymbol{v}_\alpha(z)/K_z\,_\zeta,\,\delta))\,\zeta^{\frac{1}{2}\,v_2}\,z^{\frac{1}{2}\,v_1 - 1}/(v_2\,\zeta + v_1 z)^{\frac{1}{2}\,v}.\;dz \quad (2)$$

where $c = v_1^{\frac{1}{2}v_1} v_2^{\frac{1}{2}v_2}/B(\frac{1}{2}v_1, \frac{1}{2}v_2)$ and $S_v(t, \delta)$ is the cumulative probability distribution function of the non-central Student-$t$ distribution with $v$ degrees of freedom and non-centrality parameter $\delta = (\mu_1 - \mu_2)/(\sigma_1^2/n_1 + \sigma_2^2/n_2)^{\frac{1}{2}}$, see Appendix 1.

A comparison of the powers of the statistics $T(\zeta)$ and $V$ for any specified value of $\zeta$ could be computed using the expressions (1) and (2) above (see Appendix 4).

§ 4.  <u>The Fisher-Behrens solution</u>.

The Fisher-Behrens test for the two-sample problem is also based on the statistics

$$V = (\tilde{X}_1 - \tilde{X}_2)/(S_1^2/n_1 + S_2^2/n_2)^{\frac{1}{2}} \quad \text{and} \quad Z = S_1^2/S_2^2.$$

This test is seriously flawed since the Fisher-Behrens test criteria can be obtained from an analysis in which a *confidence* calculation is treated as a *probability*.  (It is ironic that it was Fisher himself who first made the distinction between these two concepts.)

Derivation of the Fisher-Behrens criterion.

It has been shown in Lemma 2 that the conditional probability $\mathbf{Pr}\{V \le v | Z = z, \zeta\} = S_v(v/K_{Z,\zeta})$, where $S_v$ is the cumulative probability distribution function of the Student-$t$ distribution with $v$ degrees of freedom and $K_{Z,\zeta}^2 = (\zeta n_2 + n_1)(v_1 Z + v_2 \zeta)/v \zeta (n_2 Z + n_1)$. This correct result is now adjoined to an incorrect, but intuitive argument involving confidence intervals to obtain the Fisher-Behrens test criterion.

Let $f_2$ be the probability density function of the $\mathbf{F}(v_2, v_1)$ distribution, then

$$\mathbf{Pr}\{x < \zeta/Z \le x + dx\} = f_2(x)\, dx \implies \text{con}\{xz < \zeta \le (x + dx)z\} = f_2(x)dx,$$

and setting $x z = \zeta'$, $\implies x = \zeta'/z$, we see that

$$\text{con}\{\zeta' < \zeta \le \zeta' + d\zeta'\} = f_2(\zeta'/z)|dx/d\zeta'|\, d\zeta' = f_2(\zeta'/z)1/z\, d\zeta',$$

and hence the 'confidence density' of $\zeta$ is

$$f_2(\zeta/z)1/z, \quad (3)$$

where $z$ is fixed.  This implies

$$f_2(\zeta/z)d\zeta/z = 1/B(\tfrac{1}{2}v_2, \tfrac{1}{2}v_1) \cdot v_2^{\frac{1}{2}v_2 - 1} v_1^{\frac{1}{2}v_1 - 1}(\zeta/z)^{\frac{1}{2}v_2 - 1}/(v_1 + v_2\zeta/z)^{\frac{1}{2}v} \cdot d\zeta/z$$

$$= 1/B(\tfrac{1}{2}v_2, \tfrac{1}{2}v_1) \cdot (\xi)^{\frac{1}{2}v_2 - 1}(1 - \xi)^{\frac{1}{2}v_1 - 1}\, d\xi$$

on setting $\zeta/z = v_2\xi/v_1(1 - \xi)$

In a similar way we have $\mathbf{Pr}\{x < Z/\zeta \le x + dx\} = f_1(x)dx$, where $f_1$ is the probability density function of the $\mathbf{F}(v_1, v_2)$ distribution. Hence $\mathrm{con}\{z/(x + dx) \le \zeta < x\} = f_1(x)dx$, and setting $z/x = \zeta$ we get $x = z/\zeta \Rightarrow dx/d\zeta = -z/\zeta^{\,2}$, or $dx = z\, d\zeta/\zeta^2$, furthermore $dx/d\zeta = -x^2 z/z^2 = -x^2/z \Rightarrow dx/x = -x\, d\zeta/z \Rightarrow z/(x + dx) \approx z(1 - dx/x)/x = \zeta - z\, dx/x^2 = \zeta - d\zeta'$. Hence $f_1(x)dx = f_1(z/\zeta)|dx/d\zeta| \, d\zeta' = f_1(z/\zeta')z/\zeta^{\,2}. \, d\zeta'/z \Rightarrow$

$$\mathrm{con}(\zeta' - d\zeta' < \zeta < \zeta') = f_1(z/\zeta')\, z^2/\zeta^{\,2}. \, d\zeta'/z$$

Since $f_1(z/\zeta')z/\zeta'^{\,2}. \, d\zeta'/z = f_2(\zeta'/z)1/z\, d\zeta'$ the two methods produce the same result. On substituting $v_2\,\xi/v_1\,(1 - \xi)$ for $\zeta'/z$ in (1) we get

$$f_2(\zeta'/z)\, d\zeta'/z = 1/\mathrm{B}(\tfrac{1}{2}v_1, \tfrac{1}{2}v_2). \, \xi^{\tfrac{1}{2}v_2 - 1}\,(1 - \xi)^{\tfrac{1}{2}v_1 - 1}\, d\xi.$$

These results are supported by a Bayesian argument:

Let $H_i$, $i = 1, 2, ..., n$, be $n$ mutually exclusive hypotheses, with $\overset{n}{\underset{i=o}{\mathbf{U}}} H_i$ a statement that is certainly true, and let $A$ be any event.
The fundamental Bayes result is then

$$\mathbf{Pr}\{H_i \,|\, A\} = \frac{\mathbf{Pr}\{A \,/\, H_i\}\mathbf{Pr}\{H_i\}}{\Sigma_{j=1}{}^n \mathbf{Pr}\{A \,|H_j\}\mathbf{Pr}\{H_j\}}$$

In the present problem let $A = \{Z\, \varepsilon\, (z, z + dz)\}$ and let $H_i = \{\zeta_o\, \varepsilon\, (\zeta_i, \zeta_{i+1})\}$, $\zeta_i < \zeta_{i+1}$, $d\zeta_i = \zeta_{i+1} - \zeta_i$, where $\zeta_o$ is the true (unknown) value of $\sigma_1^2/\sigma_2^2$, Since

$$\mathbf{Pr}\{x < Z/\zeta < x + dx\} = f_1(x)\, dx,$$

where $f_1(x)$ is the probability density function of the $\mathbf{F}(v_1, v_2)$ distribution. It follows that $\mathbf{Pr}\{A|H\}$ in the Bayes formula can be replaced by

$$\mathbf{Pr}\{Z\, \varepsilon\, (z, z + dz)|\zeta\} = f_1(z/\zeta)dz/\zeta.$$

Now assume that the following 'law' holds: $\mathbf{Pr}\{H_i\}/\mathbf{Pr}\{H_j\} = \zeta_j\, \delta\zeta_i/\zeta_i\, \delta\zeta_j$ where the values of $\delta\zeta_i$, $i = 1, ... , n$, are sufficiently small. Then the Bayes formula leads to the following result

$$\mathbf{Pr}\{\zeta_o\, \varepsilon\, (\zeta, \zeta + d\zeta)|\, Z\, \varepsilon\, (z, z + dz)\} = \frac{f_1(z/\zeta)dz/\zeta. \, d\zeta/\zeta}{\int d\zeta'/\zeta'. \, f_1(z/\zeta')\, d\, z/\zeta'} = \frac{z\, f_1(z/\zeta)\, d\zeta/\zeta^2}{z\int f_1(z/\zeta')\, d\, z/\zeta'^{\,2}}$$

Since $z\, f_1(z/\zeta)\, d\zeta/\zeta^2 = f_2(\zeta/z)\, d\zeta/z$ (see above) it follows that

$$\mathbf{Pr}\{\zeta_o\, \varepsilon\, (\zeta, \zeta + d\zeta)\,|\, Z = z\} = \mathrm{p}(\zeta|\, Z = z)d\zeta = f_2(\zeta/z)\, d\zeta/z.$$

where $\mathrm{p}(\zeta|\, Z = z)$ can be viewed as a conditional p.d.f. Thus

$\mathrm{p}(\zeta|Z = z) = f_2(\zeta/z)d\zeta/z = 1/\mathrm{B}(\tfrac{1}{2}v_2, \, \tfrac{1}{2}v_1). \, v_2^{\tfrac{1}{2}v_2 - 1}\, v_1^{\tfrac{1}{2}v_1 - 1}(\zeta/z)^{\tfrac{1}{2}v_2 - 1}/(v_1 + v_2\zeta/z)^{\tfrac{1}{2}v}. \, d\zeta/z,$

and using the result of Lemma 2, § 2, we obtain the Fisher-Behrens test criterion, which is: if $\mu_1 = \mu_2$ and the statistic $\Theta = \theta$ then $V < v$ with (nominal) probability given by the following expression

$$1/B(\tfrac{1}{2}v_1, \tfrac{1}{2}v_2) \cdot \int_0^1 S_v(v(\ )/K(\xi, \theta))\ \xi^{\tfrac{1}{2}v_2 - 1}\ (1 - \xi)^{\tfrac{1}{2}v_1 - 1}\ d\xi\ =\ 1 - \alpha\ ,$$

where $K^2(\xi, \theta) = [v_1 \sin^2 \theta / (1 - \xi) + v_2 \cos^2\theta / \xi] / v$. To establish this requires some algebra:

$$K_{Z,\ \zeta}{}^2 =\ (v_1 Z/\zeta + v_2)\ (n_2 \zeta + n_1) / v\ (n_2 Z + n_1)$$

$$=\ [v_2 n_2 \zeta + v_1 n_2 z + v_2 n_1 + v_1 n_1 z/\zeta] / v(n_2 z + n_1)$$

But $\zeta' = \zeta/z$, hence the appropriate expression for $K^2$ is obtained from

$$(v_1/\zeta' + v_2)(n_2 z \zeta' + n_1)/v(n_2 z + n_1)\ =\ (v_1 n_2 z + v_1 n_1/\zeta' + v_2 n_2 z \zeta' + v_2 n_1)/v(n_2 z + n_1)$$

Now introduce the substitution $\zeta' = v_1 \xi / v_2 (1 - \xi)$, then

$$K^2(\xi)\ =\ [v_1 n_2 z + v_2 n_1 (1 - \xi)/\xi + v_1 n_2 z\ \xi/(1 - \xi) + v_2 n_1]/v(n_2 z + n_1)$$

$$=\ [v_1 n_2 z(1 + \xi/(1 - \xi)) + v_2 n_1((1 + \xi)/\xi + 1)]/v(n_2 z + n_1)$$

$$=\ [v_1 n_2 z/(1 - \xi) + v_2 n_1/\xi] / v(n_2 z + n_1)\ .$$

If we set $z = n_1/n_2 \tan^2 \theta$, then we obtain the expression for $K^2(\xi, \theta)$ given above:

$$K^2(\xi, \theta) = [v_1 n_2 z/(1 - \xi) + v_2 n_1/\xi] / v(n_2 z + n_1)$$

$$=\ [v_1 n_2 z/(1 - \xi) + v_2 n_1/\xi] / v(n_2 z + n_1)$$

$$=\ [v_1 n_1 \tan^2 \theta / (1 - \xi) + v_2 n_1/\xi] / v\ n_1 \sec^2\theta$$

$$=\ [v_1 \sin^2 \theta / (1 - \xi) + v_2 \cos^2\theta / \xi] / v\ . \qquad\qquad \bullet$$

Tables of the Fisher-Behrens test can be obtained by iterating $v(\ )$ in the integral expression for a chosen theta such that it has the value $1 - \alpha$.

In the compilation tables of the Fisher-Behrens criterion, it is useful, for purposes of interpolation, to tabulate the case $n_1 = \infty$. Let $z' = v_1/v_2 \cdot \xi/(1 - \xi) \Rightarrow \xi = v_2 z'/(v_1 + v_2 z')$ and $(1 - \xi) = v_1/(v_1 + v_2 z')$. Under this change of variable

$$1/B(\tfrac{1}{2}v_1, \tfrac{1}{2}v_2) \cdot \xi^{\tfrac{1}{2}v_2 - 1}(1 - \xi)^{\tfrac{1}{2}v_1 - 1}\ d\xi = 1/B(\tfrac{1}{2}v_1, \tfrac{1}{2}v_2) \cdot v_2^{\tfrac{1}{2}v_2} v_1^{\tfrac{1}{2}v_1} z'^{\tfrac{1}{2}v_2 - 1}/(v_1 + v_2 z')^{\tfrac{1}{2}v} dz'$$

and $K^2(z', \theta) = (v_1 + v_2 z')(\sin^2\theta + \cos^2\theta/z')/v$.

Therefore as $n_1 \rightarrow \infty$, $K^2(z', \theta) \rightarrow (\sin^2\theta + \cos^2\theta/z')$, and

$$1/B \cdot v_2^{\tfrac{1}{2}v_2} v_1^{\tfrac{1}{2}v_1} z'^{\tfrac{1}{2}v_2 - 1}/(v_1 + v_2 z')^{\tfrac{1}{2}v} dz' = 1/B(\tfrac{1}{2}v_1, \tfrac{1}{2}v_2) \cdot (v_2/v_1)^{\tfrac{1}{2}v_2} v_1^{\tfrac{1}{2}v} z'^{\tfrac{1}{2}v_2 - 1}/(v_1 + v_2 z')^{\tfrac{1}{2}v} dz'$$

$$= 1/B(\tfrac{1}{2}v_1, \tfrac{1}{2}v_2) \cdot (v_2/v_1)^{\tfrac{1}{2}v_2} z'^{\tfrac{1}{2}v_2 - 1}/(1 + v_2 z'/v_1)^{\tfrac{1}{2}v} dz'$$

where $(1 + v_2 z'/v_1)^{\frac{1}{2}v} = (1 + v_2 z'/v_1)^{\frac{1}{2}v_1} (1 + v_2 z'/v_1)^{\frac{1}{2}v_2} \rightarrow \exp(\frac{1}{2}v_2 z')$ as $n_1 \rightarrow \infty$ .

Therefore the tabular value of $v(\theta)$ at $\theta$ of the Fisher-Behrens test criterion for $v_2$, $v_1 = \infty$ *nominal* probability of Type-I error $= \alpha$ is that $v$ that satisfies the equation

$$(\tfrac{1}{2} v_2)^{\frac{1}{2}v_2} / \Gamma(\tfrac{1}{2}v_2) . \int_o^\infty \Phi(v/(\sin^2\theta + \cos^2\theta/z')) \; z'^{\,\frac{1}{2}v_2 - 1} \exp(-\tfrac{1}{2}v_2 z') \; dz' = 1 - \alpha .$$

The Fisher and Behrens test criteria do not satisfy condition ii in § 1, see Appendix 3, Figures 6.3 and 6.7.

§ 5. <u>The 'ideal' test for the two sample problem.</u>

In § 2 integral expressions are given for the $\mathbf{Pr}\{V < v(\theta)|\psi\}$, where $v(\theta)$ is any test criterion. If we can find a criterion $v(\theta)$ that satisfies the integral equation

$$\mathbf{Pr}\{V \leq v|\psi\} = 1/B(\tfrac{1}{2} v_1, \tfrac{1}{2} v_2) . \int_o^1 S_v \left( v(\theta)/K_z \, \dot{\wr} \right) x^{\frac{1}{2}v_1 - 1}(1 - x)^{\frac{1}{2}v_2 - 1} \, dx \text{ , (see § 2),}$$
$$= 1 - \alpha \text{ for all } \psi,$$

then $v(\theta)$ will be the 'ideal' solution. An algorithm was devised to solve this integral equation numerically for specified $\alpha$ and sample sizes $n_1$, $n_2$, which, when applied carefully to a good trial function $v(\theta)$ for $v(\theta)$, successfully computed accurate approximations to many 'ideal' criteria $v(\theta)$ , including all of those presented here.

Clearly the angle $\psi$ must be restricted to a finite representative number of values over its range. The simple lattice of $\psi$ values 0°, 1°, 2°, ... , 89°, 90° was replaced by a lattice of $\psi$ values given by $\psi = \arctan((\varphi - 45°)/45°) . + 45°$, $\varphi = 0°, 1°, 2°, ... , 90°$ . The following table gives some of the $\psi$ values in this new lattice

| $\varphi°$ | 1 | 2 | 5 | 10 | 15 | 20 | 25 | 30 | 35 | 40 | 45 |
|---|---|---|---|---|---|---|---|---|---|---|---|
| $\psi° =$ | 0.635 | 1.302 | 3.367 | 7.125 | 11.310 | 15.945 | 21.038 | 26.566 | 32.477 | 38.660 | 45 |

| $\varphi°$ | 89 | 88 | 85 | 80 | 75 | 70 | 65 | 60 | 55 | 50 |
|---|---|---|---|---|---|---|---|---|---|---|
| $\psi°$ | 89.365 | 88.698 | 86.633 | 82.875 | 78.690 | 74.055 | 68.962 | 63.434 | 57.523 | 51.340 |

Thus in the interval $[3.367, 7.125[$ of the $\psi$ lattice, of length 3.758, there are 5 points of the $\varphi$ lattice Letting the $\theta$ lattice be the same as the $\psi$ lattice then the lattice of $\theta$s has an enhanced ability to represent a solution of the integral equation in the intervals (0°, 15.945°) and (74.055°, 90°) . Also convergence was improved as it propagated away from the known solutions at $\theta = 0°$ and 90°.

To start the process of iteration in a particular case, an initial estimate of the solution function of the integral equation was required. The Welch-Aspin tables were used when available, but in most cases the initial trial was a guess conforming to the criteria (i) and (ii) in §1. Accurate values of $t_v(\alpha)$ for $v_2$ and $v_1$ were provided for $\theta = 0°$ and $\theta = 90°$ respectively; these values were not subjected to the iteration process. It was found useful to 'smooth' the initial estimate of $v(\theta)$ by multiple application of the operation

$$\mathbf{S}(v(\theta_i)) = \tfrac{1}{3}\left(v(\theta_{i-1}) + v(\theta_i) + v(\theta_{i+1})\right), \ i = 1, 2, \ \dots, 89 .$$

Let $v(\ )$ be an approximation to the function $\mathbf{v}(\ )$, by taking the $x$-axis to represent $v(\ )$ at $\theta = \theta_i$ and the $y$-axis to represent $\mathbf{Pr}(V< v(\theta)|\psi_i = \theta_i) - (1 - \alpha)$. In this context $x$ is the 'cause' and $y$ is the 'effect', then for 'causes' $x_{i-} = v_-(\theta_i)$ and $x_i = v(\theta_i)$ there exist the 'effects'

$$y_{i-} = \mathbf{Pr}(V< v_-(\theta)|\psi_i = \theta_i) - (1 - \alpha) \ \text{ and } \ y_i = \mathbf{Pr}(V< v(\theta)|\psi_i = \theta_i) - (1 - \alpha) .$$

Assuming $y = s\,x + c$ the 'effect' $y$ will be 0 when the 'cause' $x$ satisfies $s\,x = -\,c$, or $x = -\,c/s$, here $s = (\,yi- - yi)/(xi- - xi\,)$ and $c = (xi-\ yi - xi\ yi-)\,/\,(xi- - xi\ ) => -\,c/s = -(xi-\ yi - xi\ yi-)\,/(\ yi- - yi)$. The following algorithm follows from these considerations. (Cf. the "method of false position".)

$v_+(\theta_i) =$

$$\frac{v(\theta_i)(\ \mathbf{Pr}(V< v_-(\theta)|\psi_i = \theta_i) - (1 - \alpha)) - v_-(\theta_i)(\mathbf{Pr}(V< v(\theta)|\psi_i = \theta_i) - (1 - \alpha))}{\mathbf{Pr}(V< v_-(\theta)|\psi_i = \theta_i) \ - \ \mathbf{Pr}(V< v(\theta)|\psi_i = \theta_i)} ,$$

where $v_-(\theta_i)$ is the previous estimate of $\mathbf{v}(\theta_i)$, $v(\theta_i)$ is its current estimate and $v_+(\theta_i)$ is the next estimate. Notice that if $v(\theta) = \mathbf{v}(\theta)$ in this algorithm we have $v_+(\theta i) = \mathbf{v}(\theta_i)$, since $\mathbf{Pr}(V< \mathbf{v}(\theta)|\psi_i = \theta_i) - (1 - \alpha) = 0$.

Initially $v(\theta_i)$ will be an approximation to $\mathbf{v}(\theta_i)$, $i = 1, 2, \dots, 89$, and $v_-(\theta_i) = v(\theta_i) - \delta$, where $\delta$ should be the estimated average accuracy of $v(\theta_i)$, $i = 1, 2, \dots, 89$. Thus substituting the components of the vectors

$$v_- = \ v_-(\theta_1) \quad v_-(\theta_2) \quad v_-(\theta_3) \ \dots \ v_-(\theta_{88}) \quad v_-(\theta_{89})$$

$$v = \ v(\theta_1) \quad\ v(\theta_2) \quad\ v(\theta_3) \ \dots \ v(\theta_{88}) \quad\ v(\theta_{89})$$

into the algorithm yields the new vector

$$v_+ = \ v_+(\theta_1) \ \ v_+(\theta_2) \ v_+(\theta_3) \dots \ v_+(\theta_{88}) \ v_+(\theta_{89}) .$$

By replacing the vectors $v_-$ with $v$, then $v$ with $v_+$, the first step of an iterative process is established. This iterative process could be terminated either when $|\mathbf{Pr}(V< v(\theta)|\psi_i = \theta_i) - (1 - \alpha)|$ is sufficiently small for each $i = 1, 2, \dots, 89$, or the difference in the values of $v(\theta_i)$ in successive iterations for each $i = 1, 2, \dots, 89$, is less than, say, $0.00005$ .

For the cases when $n_1 \rightarrow \infty$ with $n_2$ finite we start from a result in § 2: the required criteria will be the solutions of the integral equation

$$\mathbf{Pr}\{V \leq \mathbf{v}_c(C)|\ \gamma\} = (\tfrac{1}{2} v_2)^{\tfrac{1}{2} v_2}/\Gamma(\tfrac{1}{2} v_2) \cdot \int_o^\infty \Phi(t\ \mathbf{v}_c(\ \gamma\ /(1 - \gamma)t + \gamma))\ t^{\tfrac{1}{2}v_2 - 1}\ e^{-\tfrac{1}{2}v_2 t}\ dt$$

$$= 1 - \alpha \ \text{ for all } \ \gamma .$$

This integral equation was solved using the algorithm and iteration method described above and lattices of $c$ and $\gamma$ consisting of 101 equidistant points in the interval [0, 1]. (Since $0 \le \gamma \le 1$, the function $\gamma /(1 - \gamma)t + \gamma$ is in the interval [0, 1] for all $t \ge 0$.)
Another possibility is to use $v_c(c_i) = v_c(\sin^2\theta_i) = v_\theta(\theta_i)$ together with the chosen
lattice $\theta_i$, $i = 0, 1, 2, \ldots, 90$, where polynomial interpolation of the function $v_\theta(\theta_i)$, $i =
0, 1, \ldots, 90$, is determined by the smallest $i$ : $\sin^2\theta_i \ge \gamma_j /(1 - \gamma_j)t + \gamma_j \ge 0$, where $\gamma_j =
\sin^2\psi_j$, $j = 1, 2, \ldots, 89$. Then the solutions of the integral equation for $v_1 \to \infty$ will be
directly comparable to those obtained using the method described above for finite $v_1$.

Table 5.1, Appendix 3, shows the extent to which convergence to $\boldsymbol{v}_\alpha(\theta)$ was achieved
by these programs, where, e.g., the first entry in this Table, namely $0.0^4139$, is an
abbreviation of 0.0000139. In general it was easier to obtain good convergence in
cases where either the sample sizes were both not too small ($n_1 > 10$, $n_2 > 10$), and $\alpha$
is not too small ($\alpha \ge 0.025$). In general convergence was weaker and slower for 'ideal'
test criteria which had fluctuations or irregularities.

Figures 5.1 and 5.2, Appendix 3, show 'ideal' criteria with irregularities that would
be awkward to tabulate accurately. Figure 5.3 shows the 'ideal' test criteria at the
indicated significance levels for $(v_1, v_2) = (10, 10)$, $(10, 15)$ and $(15, 15)$. When the
sample sizes are both greater than 10 and $\alpha \ge 0.025$ and when the sample sizes are
both greater than 15 and $\alpha \ge 0.005$ the irregularities in the 'ideal' criteria vanish
allowing an accurate tabulation of critical values of $V$. Table 5.2, Appendix 3,
presents irregularity-free test criteria to three decimal places for $\alpha = 0.025$. In these
tables the values of $v_2$ and $v_1$ were chosen to facilitate interpolation. (Under the
transformation $30/v$ these $v$, namely 10, 15, 30, $\infty$, transform to the integers 3, 2, 1, 0.)

The Welch-Aspin test criteria for the two sample problem are presented in Biometrika
Tables for Statisticians, Volume 1, Table 11, p. 135. The only entry in the Welch-
Aspin tabulation that seems to be in question is the entry 1.74 for $v_2 = 6$ , $v_1 = 6$, $\alpha =
0.05$, $c = 0.5$, which should be 1.73 according to Figure 5.2.

§ 6. Simulations.

All the simulations presented here conform to a common description. For each series
of simulations the sample sizes $n_1$ and $n_2$ and the test criteria of the test under scrutiny
(either the Fisher-Behrens or the 'ideal' test) were computed for these sample sizes
for each of $1 - 2\alpha = 0.1, 0.2, 0.3, \ldots$, (up to 0.9 and 0.95 if possible) , see Figures 6.3
and 6.7 . For each simulation a value $\zeta = \sigma_1^2 / \sigma_2^2$ or $\psi$ was chosen in advance and,
using a random generator, a random sample of size $n_1$ constructed on the standardized
normal distribution $\mathbf{N}(0, \zeta)$ and another random sample of size $n_2$ constructed on the
normal distribution $\mathbf{N}(0,1)$. Subsequently the statistics $V$ and $Z$ were computed for
these random samples and this statistical pair was referred to the test criteria described
above.

In the case of a simulation of the performance of the Fisher-Behrens test , see Appendix 3, Figures 6.1 and 6.2, a "probability", or confidence, $1 - 2\alpha$ was assigned to each simulated point $|V|$, $\Theta(Z)$ where $\alpha$ was obtained by evaluating

$$1/B(\tfrac{1}{2}v_1, \tfrac{1}{2}v_2) . \int_o^1 S_v(|V|/K(\xi, \Theta(Z))) \; \xi^{\frac{1}{2}v_2 - 1} \, (1 - \xi)^{\frac{1}{2}v_1 - 1} \, d\xi \;=\; 1 - \alpha \, .$$

For simulations concerned with 'ideal' criteria, critical values $x_i$ for significance levels $1 - 2\alpha = 0.0, 0.1, 0.2, 0.3, \ldots$ (to $0.7$ when $n_1, n_2 = 3, 3$ , and up to $0.9$ when $n_1, n_2 = 6, 6$, see Appendix 3, Figures 6.3 and 6.7) were assigned to each $\Theta(Z)$ of a simulation $|V|$, $\Theta(Z)$ by interpolating the critical values at the adjacent tabular values of $\theta$ .

Since the critical values $x_i$ are not equally spaced, Lagrange polynomial expressions are required for interpolation/extrapolation between values of $x_i$ There are three quadratic Lagrange polynomials associated with $x_{i+2} > x_{i+1} > x_i$ , namely those polynomials that are equal to 1 at one of these points and is equal to 0 at the other two points : let $L(x , x_{i+2} > x_{i+1} > \underline{x}_i)$ be the Lagrange polynomial that is equal to 1 at $x_i$ and equal to 0 at $x = x_{i+2}$ and $x_{i+1}$ , and $L(x , x_{i+2} > \underline{x}_{i+1} > x_i)$ equal to 1 at $x_{i+1}$ and equal to 0 at $x = x_{i+2}$ and $x_i$, with $L(x , \underline{x}_{i+2} > x_{i+1} > x_i)$ similarly defined. The most interesting case is when the function $G(x)$ was approximated by a quadratic in $x$ by using

$$0.7 \; L(x: \underline{x}_7 > x_6 > x_5) + \; 0.6 \; L(x: x_7 > \underline{x}_6 > x_5) + 0.5 \; L(x: x_7 > x_6 > \underline{x}_5) \;=$$

$$0.7 \, (x - x_6)( \; x - x_5)/(x_7 - x_5)(x_7 - x_6) + 0.6 \, (x - x_7)( \; x - x_5)/(x_6 - x_5)(x_6 - x_7) +$$
$$+ \; 0.5 \, (x - x_7)( \; x - x_6)/(x_5 - x_7)(x_5 - x_6) \, . \quad (1)$$

To attribute a probability $1 - 2\alpha$ to the simulation $|V|$ over the ranges $0.55 - 0.70$ (interpolation) and $0.70 - 0.85$ (extrapolation) set $x = |V|$ in (1) In all other cases interpolation was carried out by choosing $x_{i+2} > x_{i+1} > x_i$ so that the simulation $|V|$ was in the interval $(x_{i+2} , x_i)$.

In the case $n_1, n_2 = 6, 6$ with $|V| > 1$, *inverse* extrapolation carried out using

$$0.9 \; L(x: \underline{x}_9 > x_8 > x_7) + \; 0.8 \; L(x: x_9 > \underline{x}_8 > x_7) + 0.7 \; L(x: x_9 > x_8 > \underline{x}_7)$$

with each $x_i$ replaced by $1/x_i$ and $|V|$ by $1/|V|$ . .

For each simulation the relevant process, as explained above, was applied, after which the particular sequence of confidence levels/probabilities was obtained, $\pi_i$, $i = 1, 2, \ldots , 5000$, and this sequence was ranked to obtain the related ranked sequence $\rho_i$ , $i = 1, \ldots , 5000$ , to which were adjoined $\rho_0 = 0$ , $\rho_{5001} = 1$, thus $\rho_{i-1} \leq \rho_i$ , $i = 1, 2, \ldots , 5000$ . Finally, a graph was constructed by drawing horizontal segments between the points $x_i$, $y_i$ and $x_{i+1}$, $y_i$ , where

$$x_i = \rho_i , \; y_i = i/5000 , \; x_{i+1} = \rho_{i+1}, \, y_i = i/5000$$

for $i = 0, 1, 2, \ldots 5000$ .

(An alternative to the above procedure would have been to count, on Figures 6.3 and 6.7 the number of simulated points ($|V|$, $\Theta = \tan^{-1}(n_2 Z/n_1)^{\frac{1}{2}}$ ) falling between the criteria with $1 - 2\alpha_i = i/10$ and $1 - 2\alpha_{i+1} = (i + 1)/10$, $i = 0, 1, 2, 3, \ldots$ . If, by hypothesis, each of these events is equi-likely, the expected number for each count is 500; or 500$m$ for a multiplicity $m$ of these regions. The observed frequencies of the points ($|V|$, $\Theta$) falling into these classifications could be used to test this and other hypotheses by the application of a standard $\chi^2$ tests. This procedure was not carried out since the adopted method was deemed to be statistically more powerful.)

Simulations concerning Fisher-Behrens criteria.

Figure 6.1, Appendix 3, shows the results of 5000 simulations of $|V|$, $\Theta(Z)$ in the case $n_1 = 2$, $n_2 = 2$, $\psi = 45^o (\zeta = 1)$, together with its theoretical distribution.

Figure 6.2, Appendix 3, shows the results of 5000 simulations of $|V|$, $\Theta(Z)$ in the case $n_1 = 3$, $n_2 = 2$, $\psi = 53.5^o$ ($\zeta = 1$), together with its theoretical distribution.

The theoretical versions of the simulated distributions were computed using the formulae of §2 by substituting the Fisher-Behrens test criterion $v_\alpha(\theta)$ at each of the levels $1 - 2\alpha = 0.1, 0.2, 0.3, \ldots, 0.9$, with the same value of $\psi$ , or $\zeta$ , chosen for the simulation. Thus the probabilities $\mathbf{Pr}(|V| < v_\alpha(\theta)|\zeta)$ were calculated for each of these levels, giving the points $1 - 2\alpha$, $\mathbf{Pr}(|V| < v_\alpha(\theta)|\zeta)$ to which the points 0.0, 0.0 and 1.0, 1.0 were adjoined. These points were then connected (by a cubic spline) to obtain a graph that accurately represents the required theoretical distribution. ( Such calculations were unnecessary in the case of the 'ideal' test criteria, see below, since the theoretical distributions, by definition, are all the same straight diagonal line. )

Simulations concerning 'ideal' test criteria,

Figure 6.3, Appendix 3, shows 'ideal' test criteria and Fisher-Behrens criteria for the case $n_1 = 3$, $n_2 = 3$.

Figure 6.4, Appendix 3, shows the results of a simulation in the case $n_1 = 3$, $n_2 = 3$, $\psi = 45^o$.

Figure 6.5, Appendix 3, shows the results of a simulation in the case $n_1 = 3$, $n_2 = 3$, $\psi = 30^o$.

Figure 6.6, Appendix 3, shows the results of a simulation in the case $n_1 = 3$, $n_2 = 3$, $\psi = 15^o$.

(In Figures 6.4, 6.5, 6.6 the extrapolation of the 'ideal' test criterion over the interval 0.7 – 0.85 was more successful than anticipated.)

Figure 6.7, Appendix 3, shows 'ideal' test criteria and Fisher-Behrens criteria for the case $n_1 = 6$, $n_2 = 6$.

Figure 6.8, Appendix 3, shows the results of a simulation in the case $n_1 = 6$ , $n_2 = 6$, $\psi = 45^o$

Figure 6.9, Appendix 3, shows the results of a simulation in the case $n_1 = 6$ , $n_2 = 6$, $\psi = 30^o$.

Figure 6.10, Appendix 3, shows the results of a simulation in the case $n_1 = 10$ , $n_2 = 10$, $\psi = 45^o$.

§ 7. <u>The Linnik phenomenon</u>.

Linnik and his team showed that the solution of two sample problem would have one strange property, namely that the critical region of, say, the statistics $|V|$ , $Z$ of size $\alpha_1$ is not necessarily a subset of the critical region of size $\alpha_2$ when $\alpha_1 < \alpha_2$ . Although the iterative procedure used to construct the Tables presented in § 5 no longer converged satisfactorily for the combinations of sample sizes and $\alpha$ , the way and the circumstances under which the this phenomenon manifests itself seem clear. Figure 5.4 shows the criteria of the 'ideal' test for sample sizes $n_1 = 2$, $n_2 = 2$ and $\alpha = 0,25, 0.3, 0.35,$ and $0.4$ . Here the best approximation to the 'ideal' test criterion for $\alpha = 0.25$ had a maximum detected imbalance in the defining integral equation of the order of $10^{-3}$, much more than was tolerated elsewhere. Despite this it is reasonable to assert that, for sample sizes $n_1 = 2$, $n_2 = 2$ , the Linnik phenomenon starts to appear for a value of $\alpha$ between $0.25$ and $0.3$ . If we call this value $\alpha_L$ then, for sample sizes $n_1 = 2$, $n_2 = 2$,. $\alpha_L \approx 0.28$ . Reference to Figure 6.3 shows that for $n_1 = 3$, $n_2 = 3$, $\alpha_L < 0.15$ . Clearly $\alpha_L$ is a function of $n_1$, $n_2$ only.

Consider the two-tailed test of $H_o : \mu_1 = \mu_2$ versus $H_1 : \mu_1 \neq \mu_2$ at the significance level $2\alpha$ when $H_o$ is true and the sample sizes $n_1$, $n_2$ are not too small (say as in the Welch-Aspin tables) so that $\alpha_L < \alpha$. For all significance levels $\alpha > \alpha_L$ the probability that the point $(|V|$ , $Z)$ lies beneath the graph of the function $\boldsymbol{v}_\alpha(z)$ (i.e. $(|V|$ , $Z) < \boldsymbol{v}_\alpha(z)$ ) is $1 - 2\alpha$ and the probability that this point will lie above the function $\boldsymbol{v}_\alpha(z)$ is $2\alpha$ , which is the basis of consistent test. However if $\alpha < \alpha_L$ anomalies arise. Referring to Figure 5.4 and assuming $\alpha_L = 0.28$ for $n_1 = 2, n_2 = 2$, then if $\alpha = 0.25$ there will be circumstances under which $H_o$ will be accepted at the $2\alpha = 0.5$ level, yet rejected at the 2x0.3 = 0.6 level: since the functions $\boldsymbol{v}_{0.25}(z)$ and $\boldsymbol{v}_{0.3}(z)$ intersect there are points $(|V|$ , $Z)$ of the sample space such that $\mathbf{Pr}\{\boldsymbol{v}_{0.3}(z) < (|V|$ , $Z) < \boldsymbol{v}_{0.25}(z)\} > 0$ . Generalizing these remarks, 'ideal' test using $V$ are not consistent for $\alpha < \alpha_L$ , whereas 'ideal' tests are consistent for all $\alpha$ such that $\alpha_L < \alpha$ , implying that conventional testing exists only at significance levels $\alpha. > \alpha_L$. All the significance levels $\alpha$ of the test criteria presented in Tables 5.2 and in the Welch-Aspin Tables clearly satisfy the inequality $\alpha_L < \alpha$ , which implies that tests using these tables will be consistent for the different significance levels of these tables.

The problem posed by the Linnik phenomenon could, perhaps, be solved in the following way. If we consider the (conditional) sample space consisting of those outcomes for which $V > \boldsymbol{v}_{\alpha L}(Z)$ , the probability of which event is $\alpha_L$ , and restricting ourselves to this new sample space, attempt to find a new function $\boldsymbol{v}'_\alpha(z)$ when $\alpha < \alpha_L$.

Assume that Ho: $\mu_1 = \mu_2$ is true. If Ho is tested at significance level $\alpha$, where $\alpha > \alpha_L$, then the test is consistent for all $\alpha' : 1 > \alpha' > \alpha$, with the probability of accepting Ho equal to $1 - \alpha$ and the probability of rejecting Ho (Type I error) equal to $\alpha$.

If, however, $\alpha < \alpha_L$ then the probability of the event $\{V < \mathbf{v}_{\alpha L}(Z)\}$ is $1 - \alpha_L$, which means that for all such $V, Z$ the null hypothesis Ho is accepted consistently for all $\alpha' : 1 > \alpha' > \alpha_L$, otherwise the event $\{V > \mathbf{v}_{\alpha L}(Z)\}$ occurs. If there exists a solution $\mathbf{v}'_\alpha(z)$ to the equation $\mathbf{Pr}\{V < \mathbf{v}'_\alpha(Z) \mid V > \mathbf{v}_{\alpha L}(Z), \zeta\} = (a - \alpha_L)/\alpha_L$ then, since $(a - \alpha_L)/\alpha_L$ is not a function of $\zeta$, implying $\mathbf{Pr}\{V < \mathbf{v}'_\alpha(Z) \mid V > \mathbf{v}_{\alpha L}(Z), \zeta\}$ functionally independent of $\zeta$, it follows that

$$\mathbf{Pr}\{V < \mathbf{v}'_\alpha(Z)\} = \mathbf{Pr}\{V < \mathbf{v}_{\alpha L}(Z)\} + \mathbf{Pr}\{V < \mathbf{v}'_\alpha(Z) \mid V > \mathbf{v}_{\alpha L}(Z)\}\mathbf{Pr}\{V > \mathbf{v}_{\alpha L}(Z)\}$$

$$= (1 - \alpha_L) + (\alpha_L - \alpha)/\alpha_L. \quad \alpha_L = 1 - \alpha.$$

The function $\mathbf{v}'_\alpha(z)$ should be greater than $\mathbf{v}_{\alpha L}(z)$, less than $\mathbf{v}_\alpha(z)$ for 'most' $z$, and should have the property (i), § 1. If for all $\alpha'$, $\alpha'' : \alpha_L > \alpha' > \alpha'' > \alpha$ the test criteria satisfy $\mathbf{v}'_\alpha(z) < \mathbf{v}'_{\alpha'}(z)$, then these criteria will be consistent, implying that consistent tests with the similarity property exist for $\alpha' : 1 > \alpha' > \alpha$.

<u>Appendix 1.</u>

<u>The $\chi^2$ distribution</u>

Let the random variable $X$ have the standardized normal distribution, *i.e.* $X \sim \mathbf{N}(0, 1)$, and let $X_i$, $i = 1, \ldots, v$, be a random sample of $v$ independent observations on $X$. Then the random variable $Y = X_1{}^2 + X_2{}^2 + \ldots + X_v{}^2$ has the $\chi^2$ distribution with $v$ degrees of freedom . The probability density function of this distribution is

$$f(y) = (½)/\Gamma(½\,v) \bullet\ y^{½\,v-1}\ \mathrm{e}^{-½\,y}\ .$$

<u>The Student-*t* distribution.</u>

The (central) Student-*t* distribution with $v$ degrees of freedom has the probability density function

$$1\ /\ \mathrm{B}(½,\ ½\,v)\ \ v^{½}(1 - t^2/v)^{½(v+1)}\ ,$$

with the cumulative probability distribution function

$$\boldsymbol{S_v}(x)\ =\ 1/\mathrm{B}(½,\ ½\,v) \bullet \int_{-\infty}^{x}\ \ dt\ /\ v^{½}(1 - t^2/v)^{½(v+1)}\ .$$

The non-central Student-*t* distribution with $v$ degrees of freedom has the probability density function

$$1\ /(\pi)^{½}\ \Gamma(½\,v)\ v^{½}\ (1 + t^2/v)^{½(v+1)} \bullet\ 2\exp(-\delta^2/2)\int_{o}^{\infty}\exp\{-u^2 + 2^{½}\,\delta\,u\,t/(t+v)^{½}\}\ u^v\ du\ .$$

See C. R. Rao: Linear Statistical Inference and its Applications, 2[nd] Edition (Wiley), p.138.

<u>The $\mathbf{F}(v_1, v_2)$ distribution.</u>

If the random variables $U_1$ and $U_2$ are independent and have $\chi^2$ distributions with respective degrees of freedom $v_1$ and $v_2$ then the probability density function of the random variable $F = v_2/v_1 \bullet U_1/U_2$ is

$$1/\mathrm{B}(½\,v_1, ½\,v_2) \bullet\ (v_1\,v_2)^{½\,v}\,f^{\,½\,v_1 - 1}\,/\,(1 + f\,v_1/v_2)^{½\,v}\ ,$$

where $v = v_1 + v_2$ . Hence the probability density function of the random variable $Z = \sigma_1{}^2 U_1/v_1 \bullet /\ \sigma_2{}^2 U_2/v_2 = v_2\sigma_1{}^2/v_1\sigma_2{}^2 \bullet U_1/U_2$ is

$$1/\mathrm{B}(½\,v_1, ½\,v_2) \bullet\ z^{½\,v_1 - 1}\zeta^{\,½\,v_2}/(v_2\zeta + v_1 z)^{½\,v}\ .$$

It can be seen that
.
$$\mathbf{E}(Z) = 1/\mathrm{B}(½\,v_1, ½\,v_2) \bullet \int_{o}^{1} z(x)\ x^{½(v_1)-1}(1-x)^{½(v_2)-1}\ dx\ ,$$

where $z = \zeta v_2 \; x / v_1(1-x)$. Hence

$\mathbf{E}(Z) = $ B($\tfrac{1}{2}(v_1+2),\tfrac{1}{2}(v_2-2))/$B($\tfrac{1}{2}v_1,\tfrac{1}{2}v_2$) $\zeta v_2 / v_1 = (v_2)/(v_2-2) \cdot \zeta$ if $v_2 \geq 3$.

Similarly $\mathbf{E}(Z^2) = $ B($\tfrac{1}{2}(v_1+4),\tfrac{1}{2}(v_2-4))/$B($\tfrac{1}{2}v_1,\tfrac{1}{2}v_2$). $(\zeta v_2 / v_1)^2$, $v_2 \geq 5$,

$$= \Gamma(\tfrac{1}{2}(v_1+4)/\Gamma(\tfrac{1}{2}v_1). \; \Gamma(\tfrac{1}{2}(v_2-4))/\Gamma(\tfrac{1}{2}v_2). \; (\zeta v_2 / v_1)^2$$

Hence $\mathbf{Var}(Z) = \mathbf{E}(Z^2) - \mathbf{E}^2(Z) =$

$$(v_1)/(v_2-2). \; [(v_1+2)/(v_2-4). - (v_1)/(v_2-2)] \; (\zeta v_2 / v_1)^2 > 0, \; v_2 \geq 5.$$

Let $Z' = 1/Z$ and $\zeta' = 1/\zeta$, then $\mathbf{E}(Z') = (v_1)/(v_1-2) \cdot \zeta'$ if $v_1 \geq 3$ and

$$\mathbf{Var}(Z') = (v_2)/(v_1-2). \; [(v_2+2)/(v_1-4). - (v_2)/(v_1-2)] \; (\zeta' v_1/v_2)^2 > 0, \; v_1 \geq 5.$$

Details of the proof of the Lemma, § 2.

We have $Z = S_1^2/S_2^2 = v_2\,\sigma_1^2\,U_1/v_1\sigma_2^2 U_2$, and set

$$W = (S_1^2/n_{1\bullet} + S_2^2/n_2) = \sigma_1^2 U_1/n_1 v_1 + \sigma_2^2 U_2/n_2 v_2 ,$$

and solving these two equations for $U_1$, $U_2$ in terms of $W$, $Z$ gives

$$U_1 = v_1 W/\sigma_1^2(1/n_1 + 1/n_2 Z) , \qquad U_2 = v_2 W/\sigma_2^2\,(Z/n_1 + 1/n_2) .$$

With $J = \partial(u_1,u_2)/\partial(w,z) = v_1 v_2(w\,(z/n_1 + 1/n_2))/\sigma_1^2\sigma_2^2(z/n_1 + 1/n_2)^3$,
and we see that the joint probability density function of $W,Z$ is

$$f_1(v_1 w/\sigma_1^2(1/n_1 + 1/n_2 z))f_2(v_2\,w/\,\sigma_2^{\,2}\,(z/n_1 + 1/n_2))\,|\partial(u_1,u_2)/\partial(w,z)|\;dw\,dz$$

$$= f_1(v_1 w/\sigma_1^2(1/n_1 + 1/n_2 z))f_2(v_2\,w/\,\sigma_2^{\,2}\,(z/n_1 + 1/n_2))$$
$$v_1 v_2 w\,/\,\sigma_1^2\sigma_2^2(z/n_1 + 1/n_2)^2\;\;dw\,dz .$$

where $f_1(\cdot)$ and $f_2(\cdot)$ are the probability density functions of the $\chi^2$ distributions with $v_1$ and $v_2$ degrees of freedom

Now consider $V = (\tilde{X}_1 - \tilde{X}_2)/W^{1/2} = X/Y^{1/2}$ where

$X = ((\tilde{X}_1 - \tilde{X}_2) - (\mu_1 - \mu_2))/(\sigma_1^2/n_{1\bullet} + \sigma_2^2/n_2)^{1/2} = ((\tilde{X}_1 - \tilde{X}_2) - (\mu_1 - \mu_2))/\sigma_2(\zeta/n_1 + 1/n_2)^{1/2}$,

implying $X \sim \mathbf{N}(\delta,1)$, $\delta = (\mu_1 - \mu_2)/(\sigma_1^2/n_{1\bullet} + \sigma_2^2/n_2)^{1/2}$,

and $Y = W/(\sigma_1^2/n_{1\bullet} + \sigma_2^2/n_2) = W/\sigma_2^2(\zeta/n_1 + 1/n_2) = W/\sigma_1^2(1/n_{1\bullet} + 1/n_2\zeta)$.

Since the functional form of the joint probability density function of $Y$, $Z$ is

$$[f_1(y\,v_1(1/n_1 + 1/n_2\zeta)\,/(1/n_1 + 1/n_2 z))f_2(y\,v_2\,(\zeta/n_1 + 1/n_2)\,/\,(z/n_1 + 1/n_2))$$
$$v_1 v_2 y\;\sigma_2^2(\zeta/n_1 + 1/n_2)\,/\,\sigma_1^2\sigma_2^2(z/n_1 + 1/n_2)^2\;\;dw\,dz$$

$$= f_1(K_1(z)\,y)f_2(K_2(z)\,y)y/(z/n_1 + 1/n_2)^2\cdot\;v_1 v_2 y\;(\zeta/n_1 + 1/n_2)\,(1/n_1 + 1/n_2\zeta)\,dy\,dz ,$$
where

$$K_1(z) = v_1(1/n_1 + 1/n_2\zeta)\,/(1/n_{1\bullet} + 1/n_2 z)\;\text{and}\;\;K_2(z) = v_2\,(\zeta/n_1 + 1/n_2)\,/\,(z/n_1 + 1/n_2) ,$$

the explicit form of $f_1(K_1(z)\,y)f_2(K_2(z)\,y)y$ is obtained by substituting the appropriate $\chi^2$ density for $f_1$ and $f_2$, which gives

$$(K_1(z)\,y)^{1/2 v_1 - 1}\;e^{-1/2 K_1(z)y}\,(K_2(z)\,y)^{1/2 v_2 - 1}\;e^{-1/2 K_2(z)y}\,y = y^{1/2(v_1 + v_2) - 1}\;e^{-1/2\,(K_1(z) + K_2(z))\,y}\;G(z).$$

It follows that the conditional probability distribution of the random variable $(K_1(Z) + K_2(Z))Y$, given $Z = z$, is the $\chi^2$ distribution, Appendix 1 with $v_1 + v_2$ degrees of freedom. ●

APPENDIX 3

Table 5.1

The departure of the test criteria in Table 5.2b from 'ideal' criteria with $\alpha = 0.025$ expressed in terms of the maximum and minimum detected differences $d = \mathbf{Pr}$(Type-I error when $\mu_1 = \mu_2$) $- \alpha$ from the 'ideal' and their associated values of $\gamma$ .(Criteria for cases when one sample sizes has $\nu = 6$ or 8 have been omitted from Table5.2b because these cases are difficult to present in tabular form, see Figures 5.1 and 5.2.)

| $\nu_2$ | $\nu_1$ | min $d$ | $\gamma_{min}$ | max $d$ | $\gamma_{max}$ | av $|d|$ |
|---|---|---|---|---|---|---|
| 6 | 6 | $-0.0^4139$ | 0.162 | $0.0^4128$ | 0.073 | $0.0^579$ |
|  | 8 | $- .0^566$ | .163 | $.0^544$ | .065 | $.0^519$ |
|  | 10 | $- .0^4133$ | .211 | $.0^4127$ | .078 | $.0^527$ |
|  | 15 | $- .0^522$ | .805 | $.0^522$ | .657 | $.0^512$ |
|  | 30 | $- .0^552$ | .417 | $.0^4242$ | .755 | $.0^571$ |
|  | ∞ | $- .0^3122$ | .780 | $.0^4430$ | .501 | $.0^4108$ |
| 8 | 8 | $-0.0^523$ | 0.091 | $0.0^512$ | 0.344 | $0.0^66$ |
|  | 10 | $- .0^539$ | .205 | $.0^522$ | .082 | $.0^68$ |
|  | 15 | $- .0^541$ | .121 | $.0^584$ | .173 | $.0^513$ |
|  | 30 | $- .0^551$ | 1.000 | $.0^4339$ | .810 | $.0^536$ |
|  | ∞ | $- .0^4297$ | .688 | $.0^4336$ | .278 | $.0^597$ |
| 10 | 10 | $-0.0^521$ | 0.116 | $0.0^518$ | 0.500 | $0.0^510$ |
|  | 15 | $- .0^546$ | .153 | $.0^565$ | .235 | $.0^511$ |
|  | 30 | $- .0^66$ | .734 | $.0^66$ | .613 | $.0^63$ |
|  | ∞ | $- .0^55$ | .861 | $.0^64$ | .778 | $.0^69$ |
| 15 | 15 | $-0.0^65$ | 0.371 | $0.0^66$ | 0.500 | $0.0^62$ |
|  | 30 | $- .0^64$ | .774 | $.0^62$ | 1.000 | $.0^62$ |
|  | ∞ | $- .0^61$ | 1.000 | $.0^63$ | .014 | $.0^61$ |
| 30 | 30 | $-0.0^64$ | 0.396 | $0.0^64$ | 0.500 | $0.0^62$ |
|  | ∞ | $- .0^62$ | .082 | $.0^62$ | .038 | $.0^61$ |
| ∞ | ∞ | 0.0 |  | 0.0 |  | 0.0 |

# Table2

**Table 2.** *The upper* $2\frac{1}{2}$ *% critical values of the statistic* $V$ *to four significant figures for* $\nu_1 \geq 10, \nu_2 \geq 10$.

| $\nu_2$ | $\nu_1$ | c = 0.00 / 1.00 | 0.05 / 0.95 | 0.10 / 0.90 | 0.15 / 0.85 | 0.20 / 0.80 | 0.25 / 0.75 | 0.30 / 0.70 | 0.35 / 0.65 | 0.40 / 0.60 | 0.45 / 0.55 | 0.50 |
|---|---|---|---|---|---|---|---|---|---|---|---|---|
| 10 | 10 | 2.228 | 2.206 | 2.185 | 2.164 | 2.143 | 2.124 | 2.108 | 2.093 | 2.077 | 2.064 | 2.059 |
|    | 15 | 2.228 | 2.206 | 2.185 | 2.163 | 2.143 | 2.123 | 2.105 | 2.088 | 2.073 | 2.059 | 2.050 |
|    | 30 | 2.228 | 2.206 | 2.184 | 2.163 | 2.142 | 2.122 | 2.103 | 2.084 | 2.067 | 2.053 | 2.040 |
|    | $\infty$ | 2.228 | 2.206 | 2.184 | 2.163 | 2.141 | 2.120 | 2.101 | 2.081 | 2.062 | 2.045 | 2.029 |
| 15 | 10 | 2.131 | 2.117 | 2.103 | 2.089 | 2.078 | 2.067 | 2.058 | 2.052 | 2.048 | 2.047 | 2.050 |
|    | 15 | 2.131 | 2.117 | 2.103 | 2.088 | 2.077 | 2.065 | 2.054 | 2.047 | 2.041 | 2.035 | 2.032 |
|    | 30 | 2.131 | 2.117 | 2.102 | 2.088 | 2.075 | 2.063 | 2.051 | 2.041 | 2.032 | 2.024 | 2.018 |
|    | $\infty$ | 2.131 | 2.116 | 2.102 | 2.087 | 2.074 | 2.060 | 2.047 | 2.035 | 2.024 | 2.013 | 2.003 |
| 30 | 10 | 2.042 | 2.035 | 2.029 | 2.024 | 2.020 | 2.018 | 2.017 | 2.019 | 2.023 | 2.030 | 2.040 |
|    | 15 | 2.042 | 2.035 | 2.028 | 2.022 | 2.018 | 2.015 | 2.011 | 2.011 | 2.012 | 2.014 | 2.018 |
|    | 30 | 2.042 | 2.035 | 2.028 | 2.022 | 2.016 | 2.011 | 2.006 | 2.003 | 2.001 | 1.999 | 1.998 |
|    | $\infty$ | 2.042 | 2.035 | 2.027 | 2.020 | 2.014 | 2.007 | 2.001 | 1.995 | 1.990 | 1.985 | 1.981 |
| $\infty$ | 10 | 1.960 | 1.960 | 1.962 | 1.965 | 1.969 | 1.975 | 1.982 | 1.991 | 2.002 | 2.015 | 2.029 |
|    | 15 | 1.960 | 1.960 | 1.961 | 1.963 | 1.966 | 1.970 | 1.974 | 1.980 | 1.987 | 1.995 | 2.003 |
|    | 30 | 1.960 | 1.960 | 1.961 | 1.962 | 1.963 | 1.965 | 1.967 | 1.970 | 1.973 | 1.977 | 1.981 |
|    | $\infty$ | 1.960 | 1.960 | 1.960 | 1.960 | 1.960 | 1.960 | 1.960 | 1.960 | 1.960 | 1.960 | 1.960 |

Table 5.2

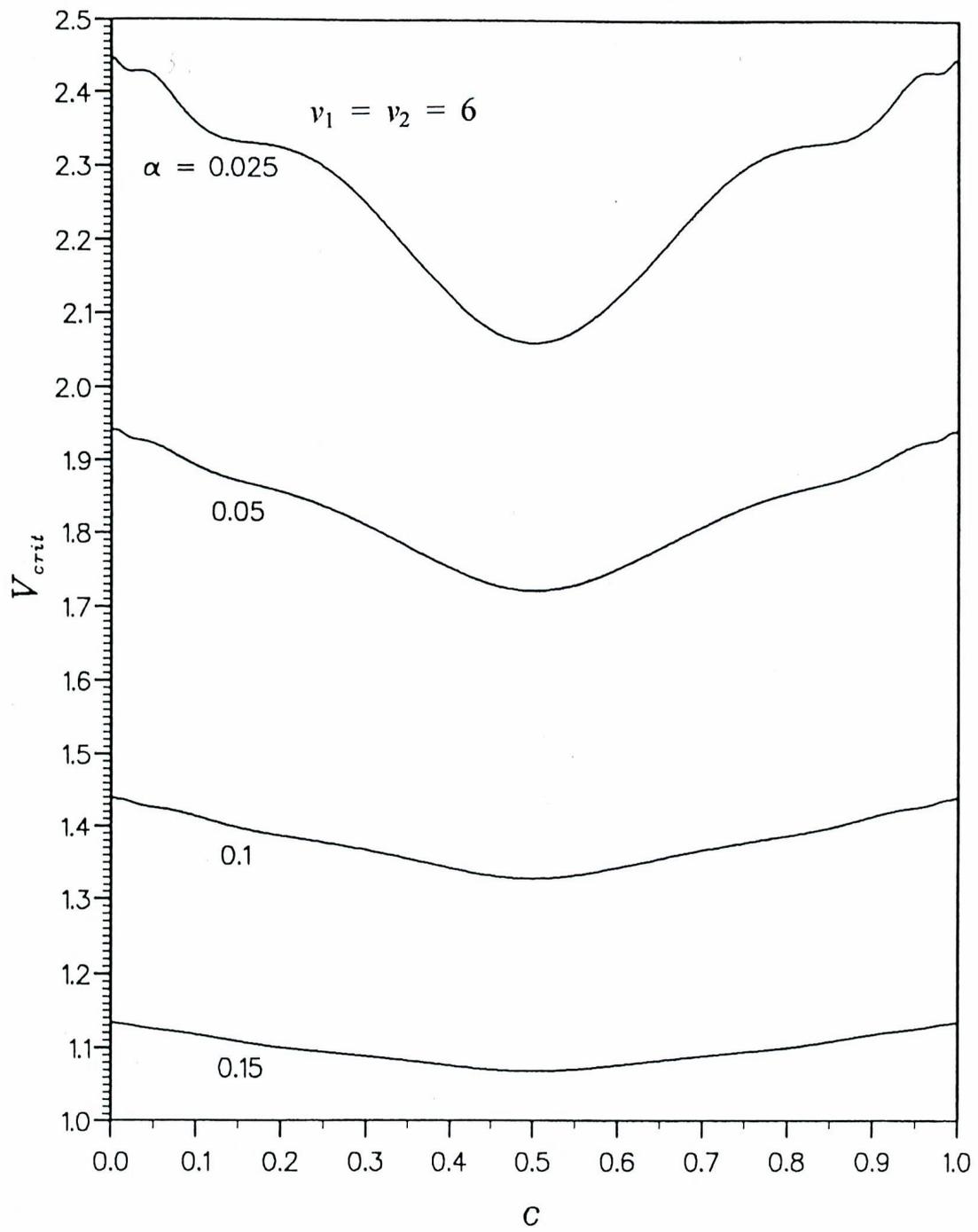

Figure 5.1

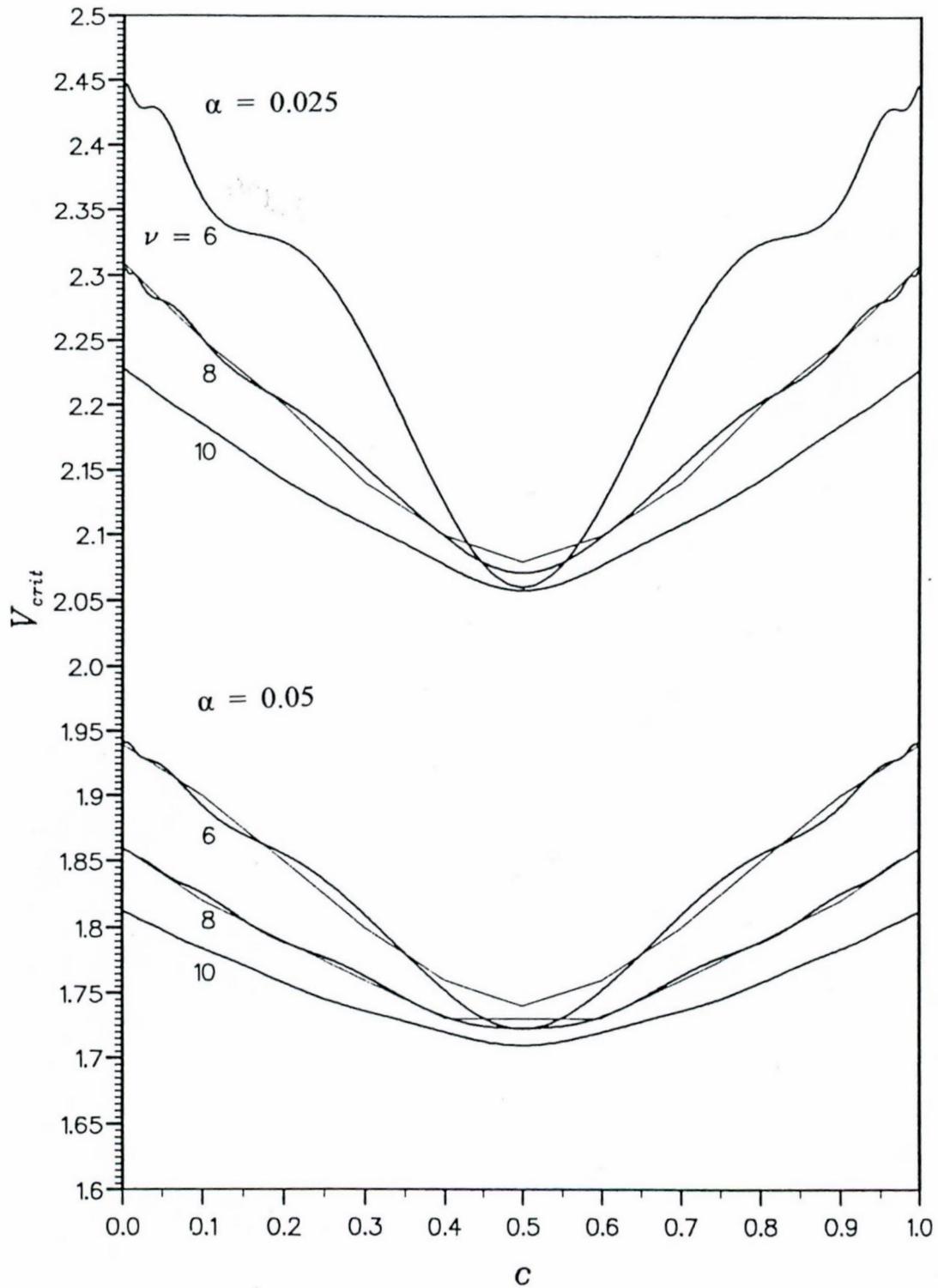

Figure 5.2

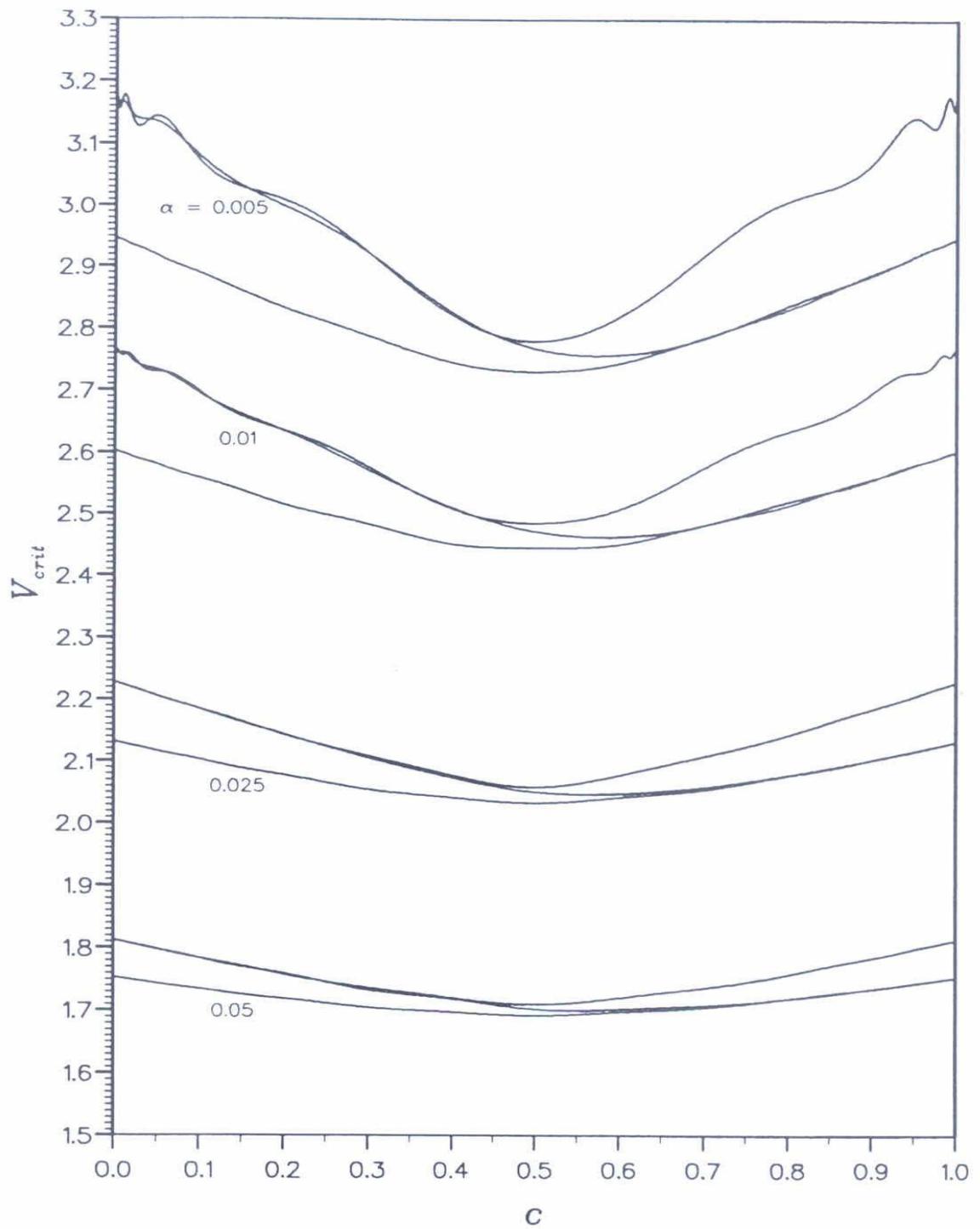

Figure 5.3

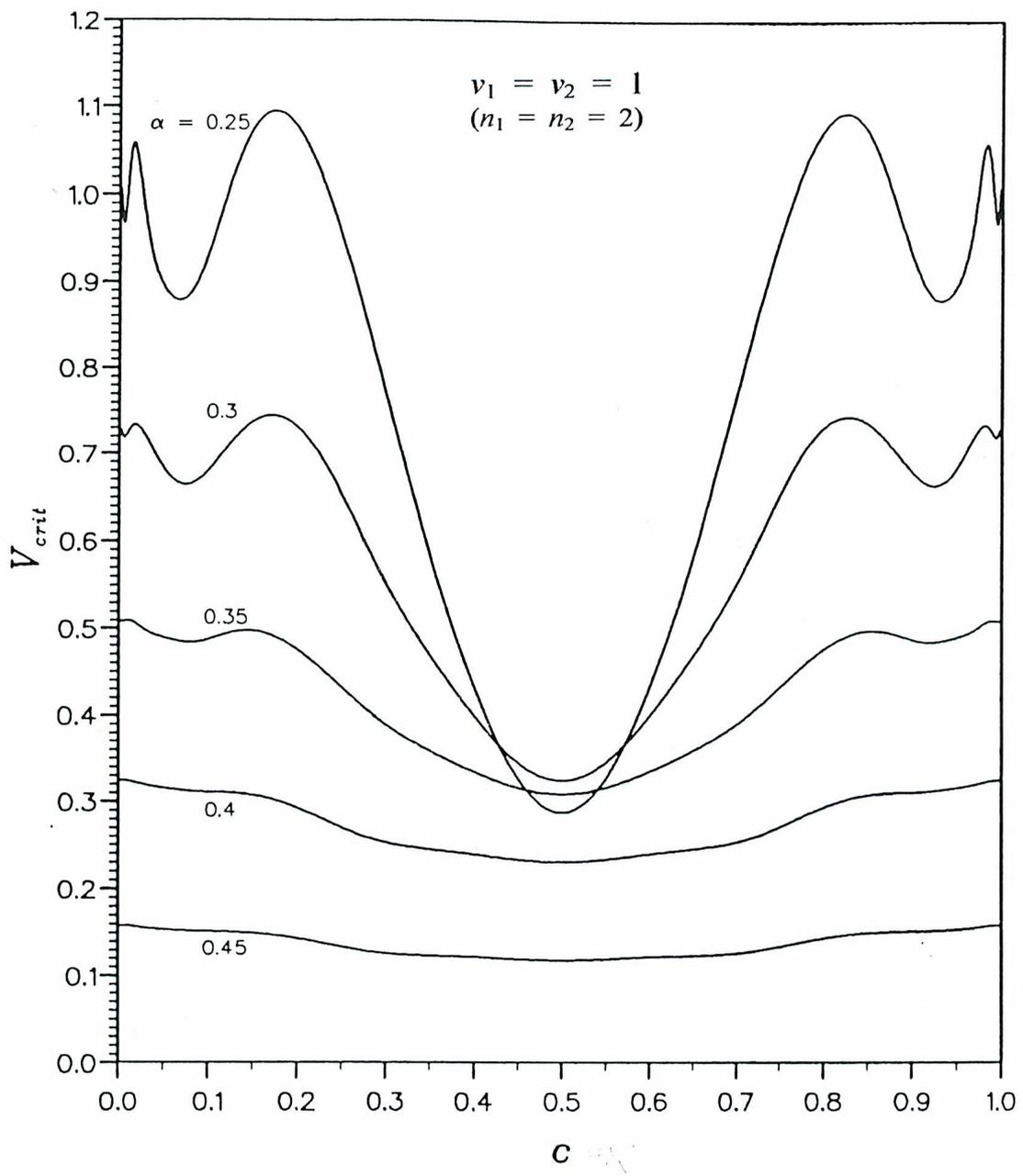

Figure 5.4

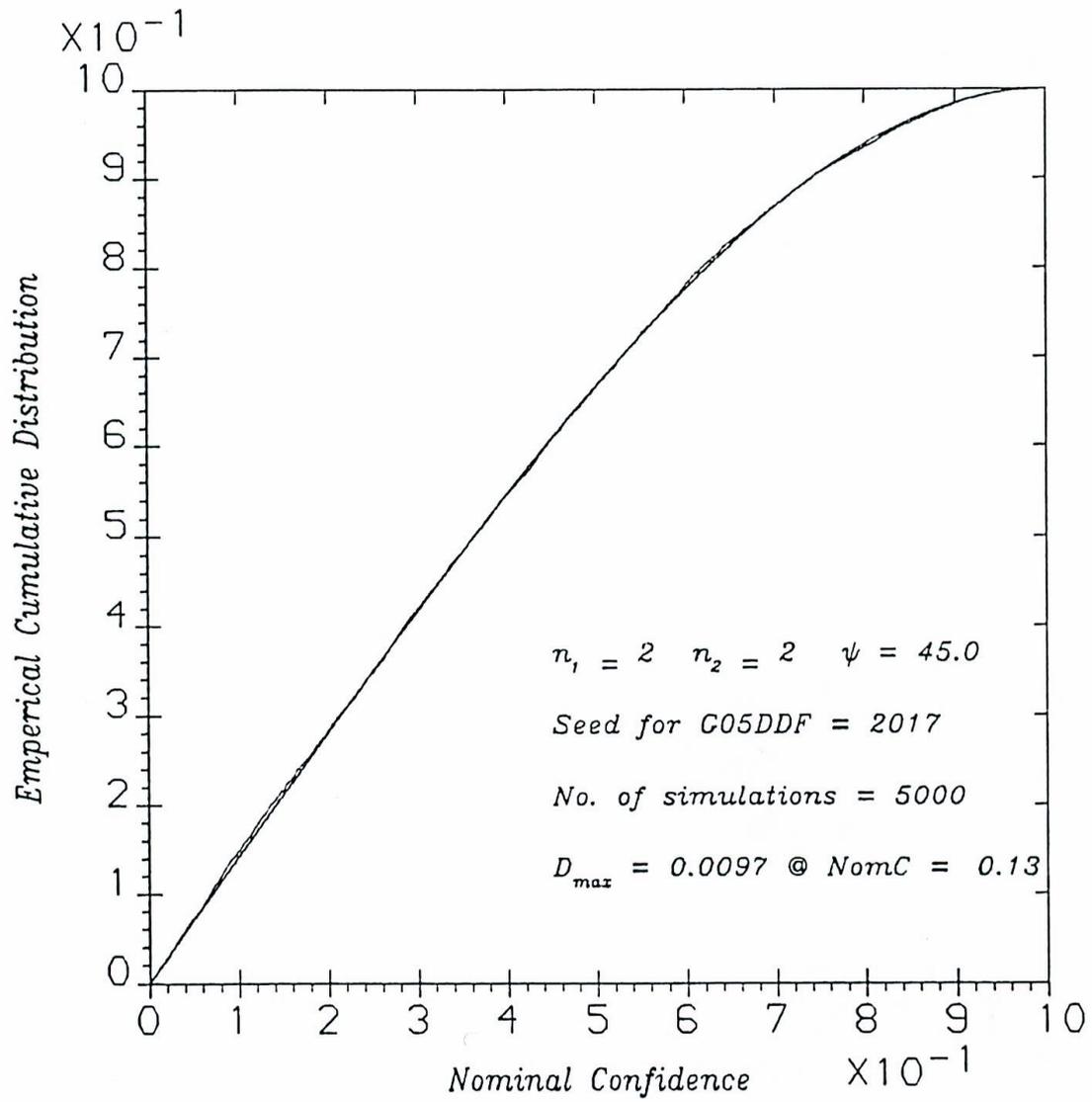

Figure 6.1

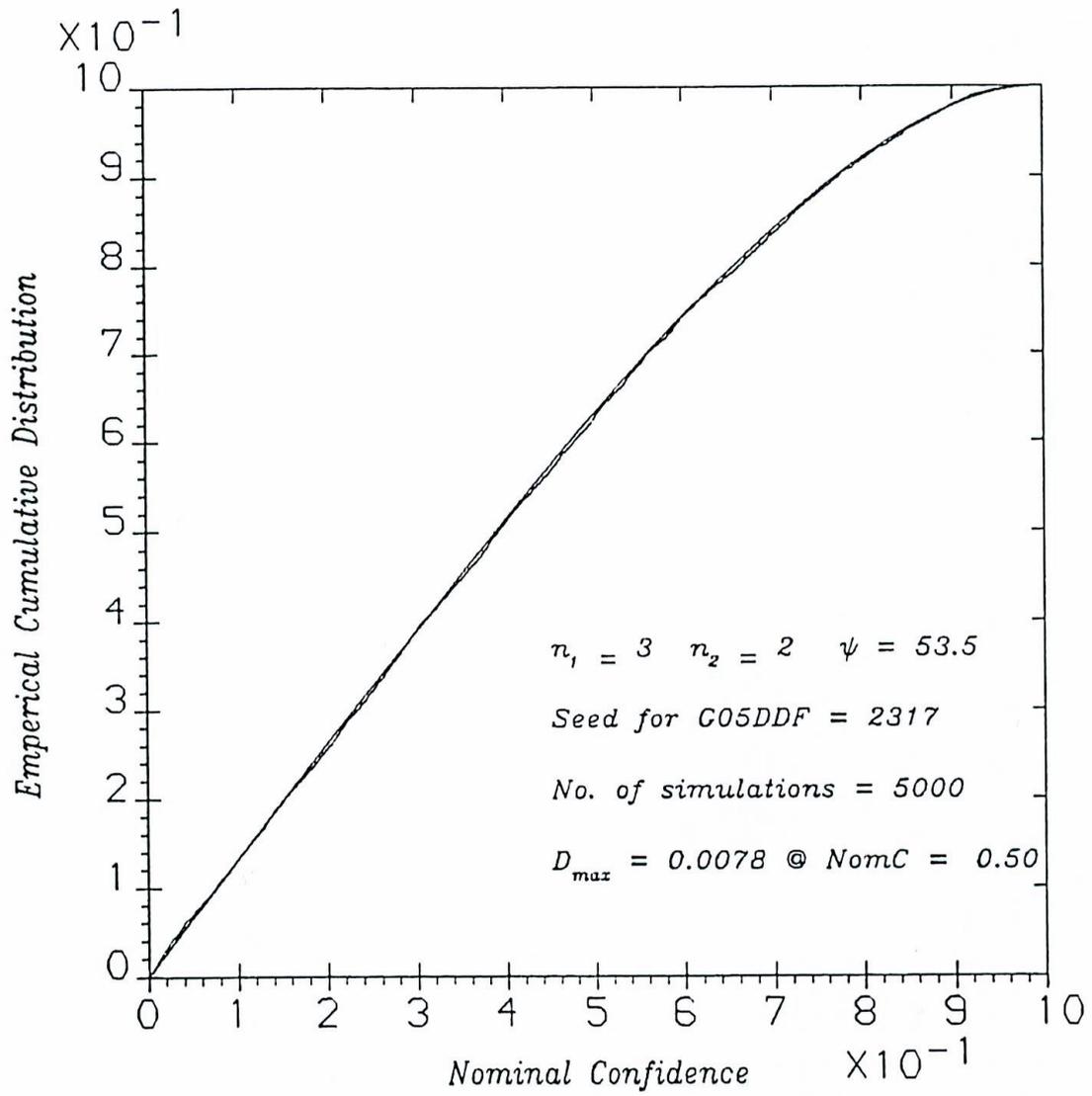

Figure 6.2

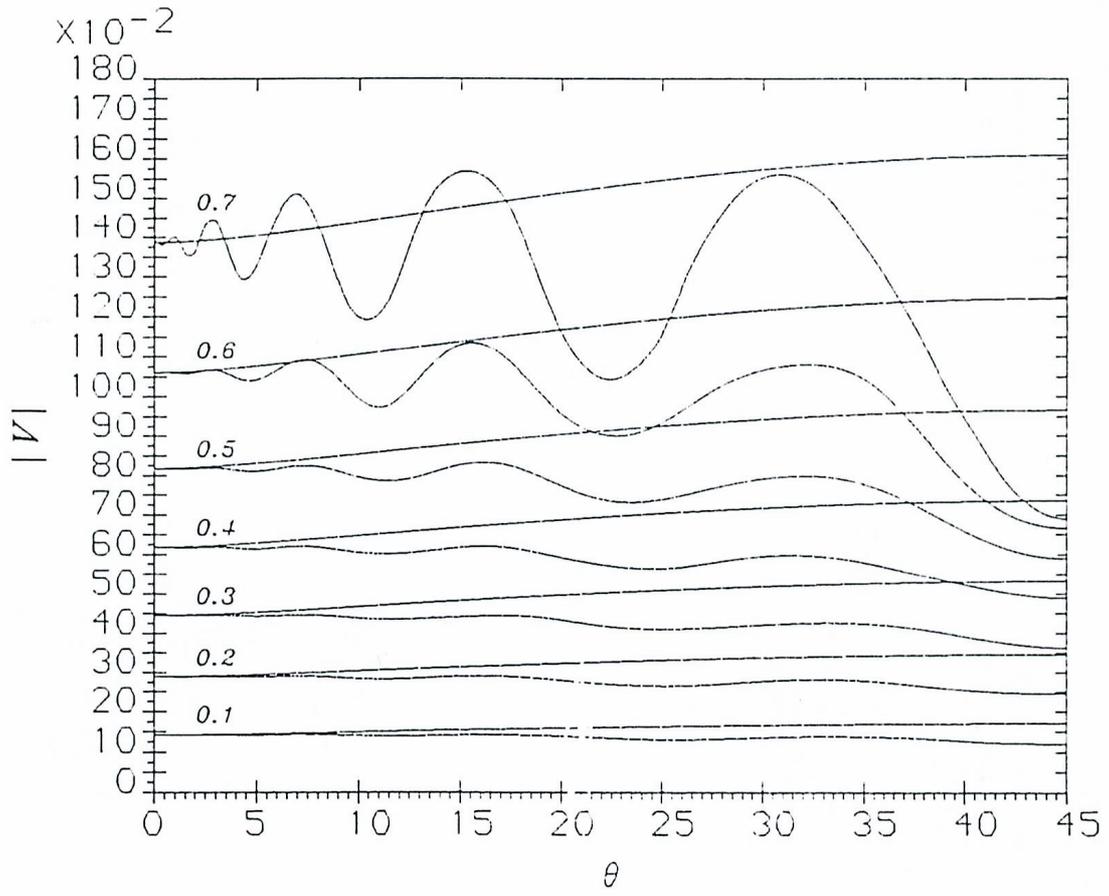

Figure 6.3

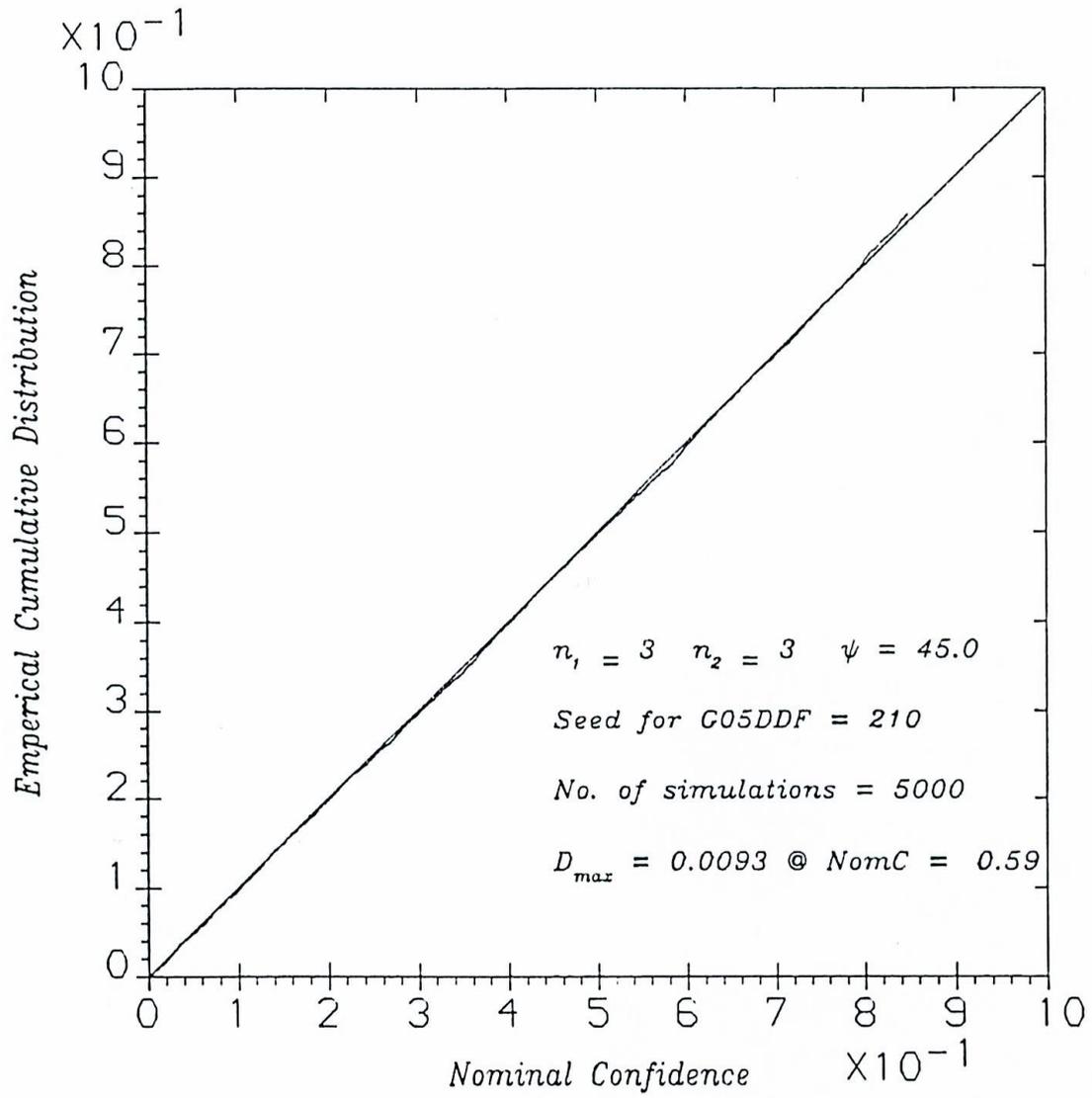

Figure 6.4

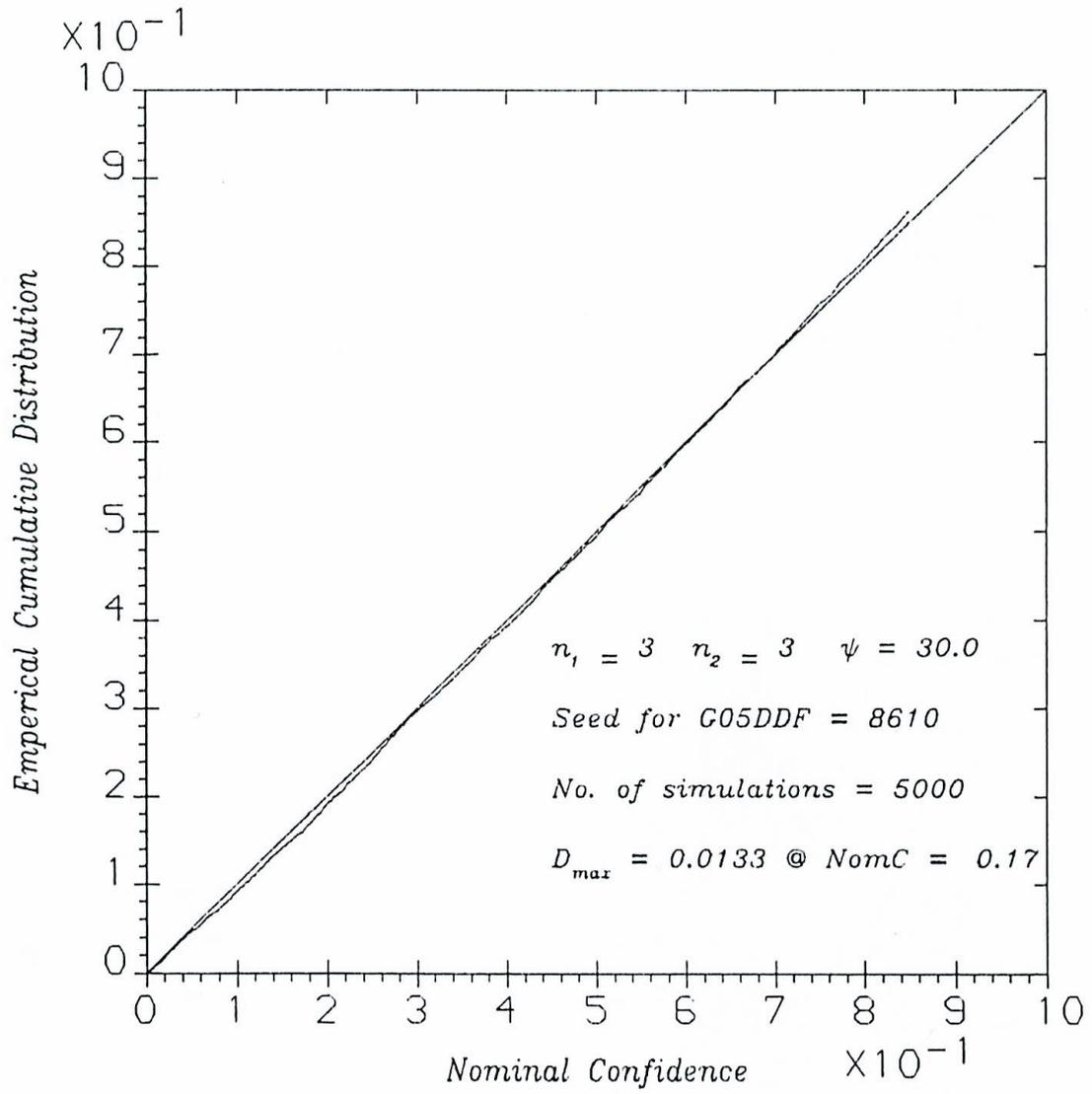

$n_1 = 3$   $n_2 = 3$   $\psi = 30.0$

Seed for G05DDF = 8610

No. of simulations = 5000

$D_{max} = 0.0133$ @ NomC = 0.17

Figure 6.5

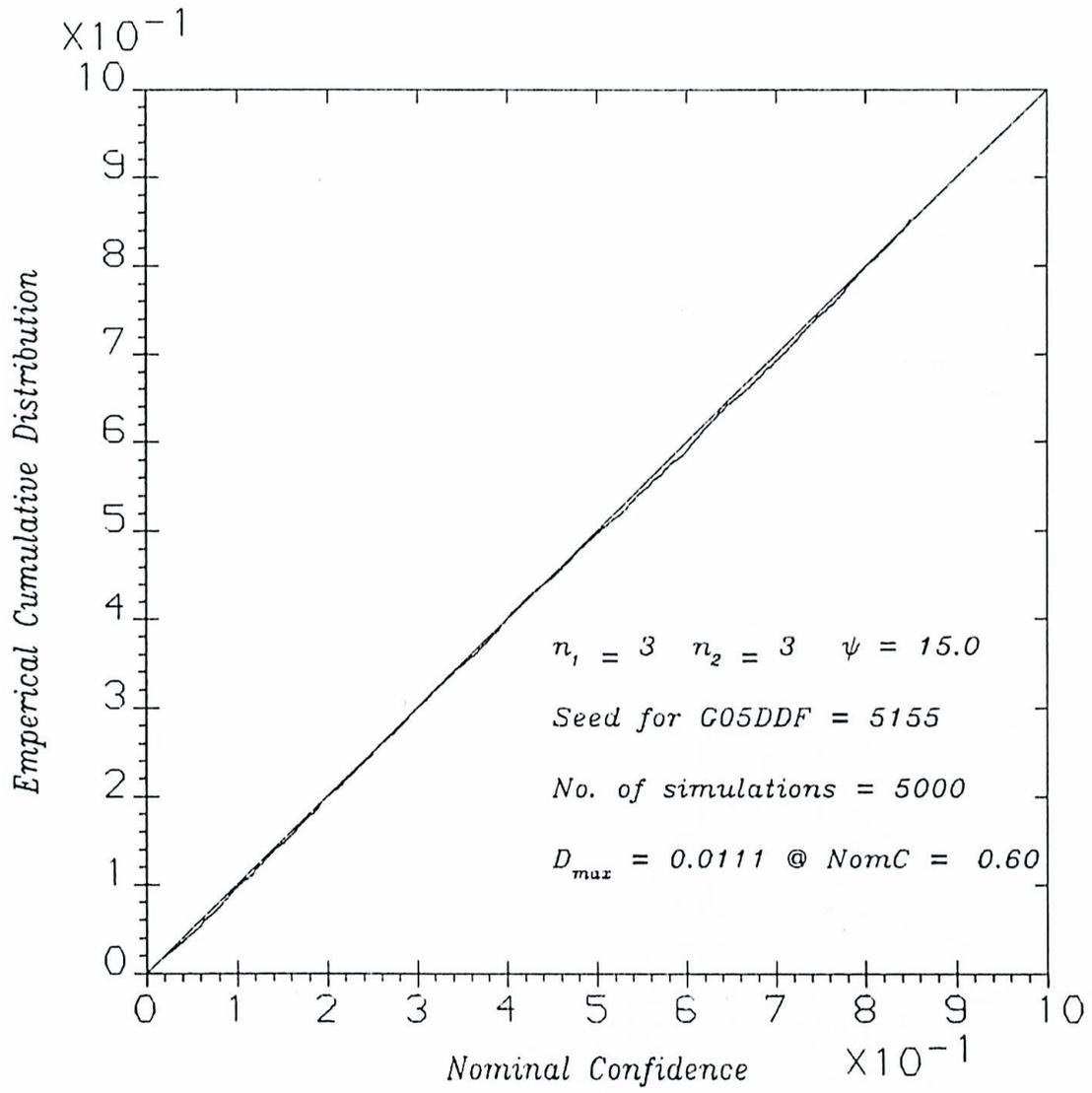

$n_1 = 3$  $n_2 = 3$  $\psi = 15.0$

Seed for G05DDF = 5155

No. of simulations = 5000

$D_{max} = 0.0111$ @ NomC = 0.60

Figure 6.6

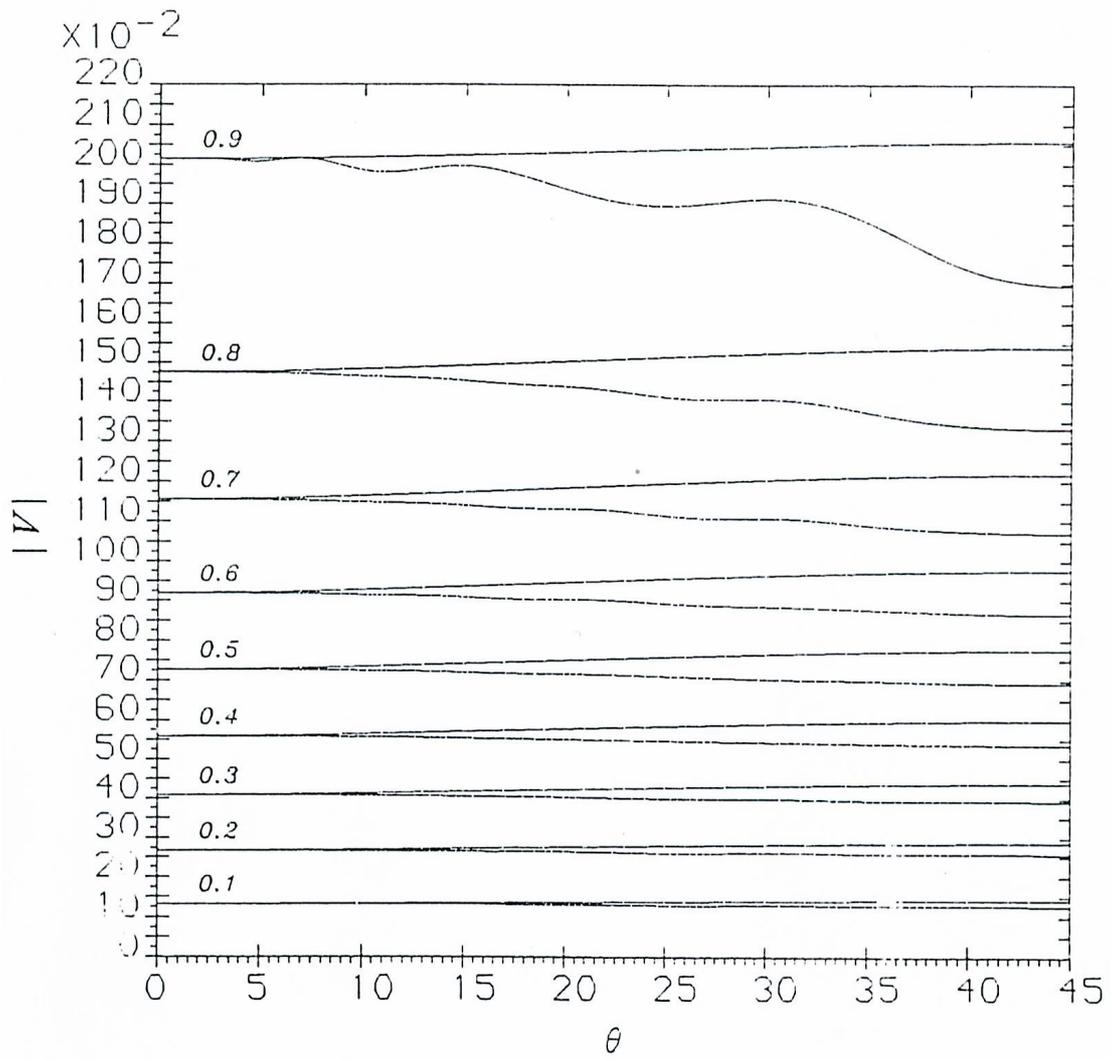

Figure 6.7

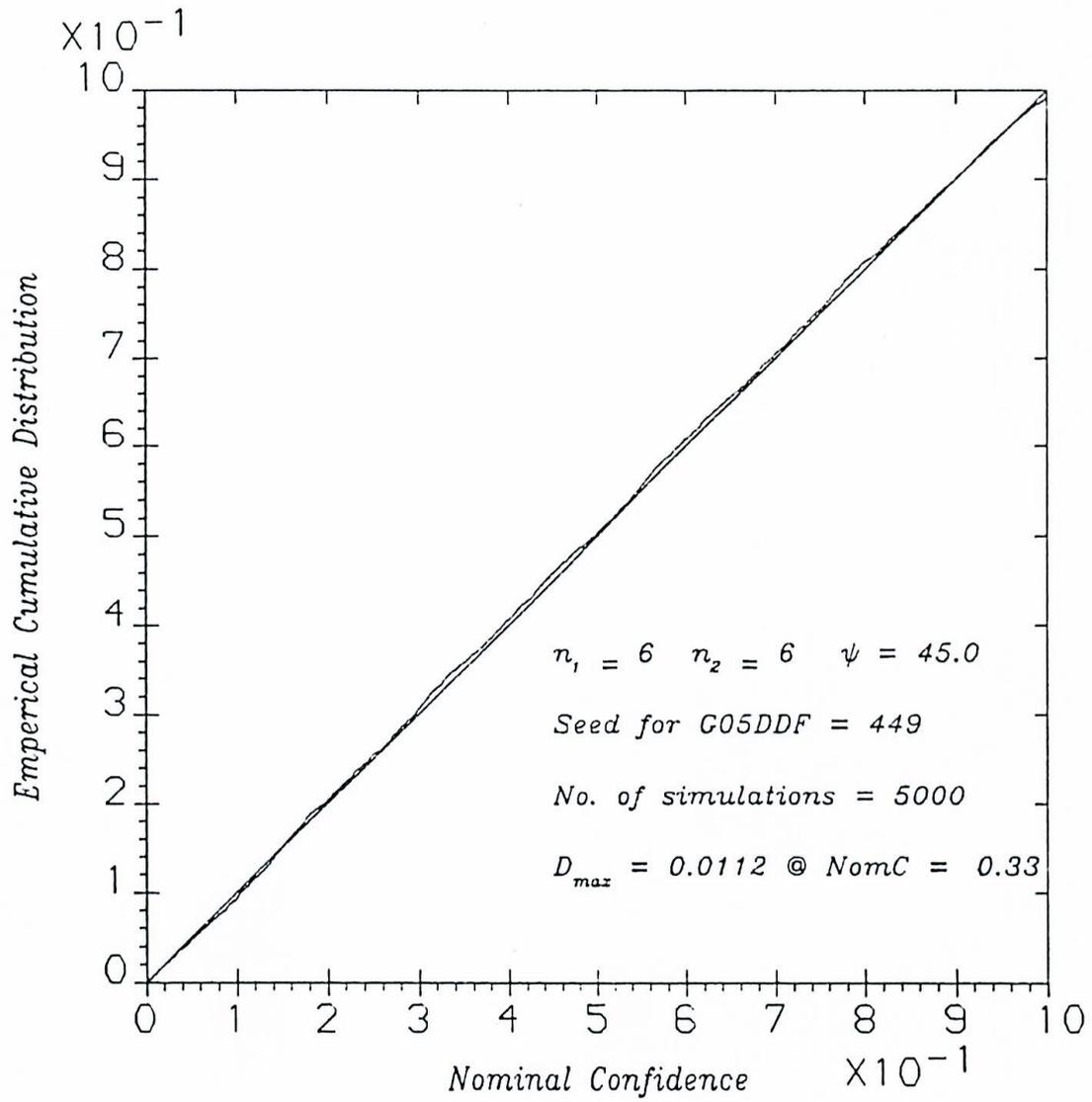

The image contains the following axis labels and annotations:

- Y-axis: $\times 10^{-1}$
- Y-axis label: *Emperical Cumulative Distribution*
- X-axis label: *Nominal Confidence* $\times 10^{-1}$

$n_1 = 6 \quad n_2 = 6 \quad \psi = 45.0$

*Seed for G05DDF = 449*

*No. of simulations = 5000*

$D_{max} = 0.0112 \ @ \ NomC = 0.33$

Figure 6.8

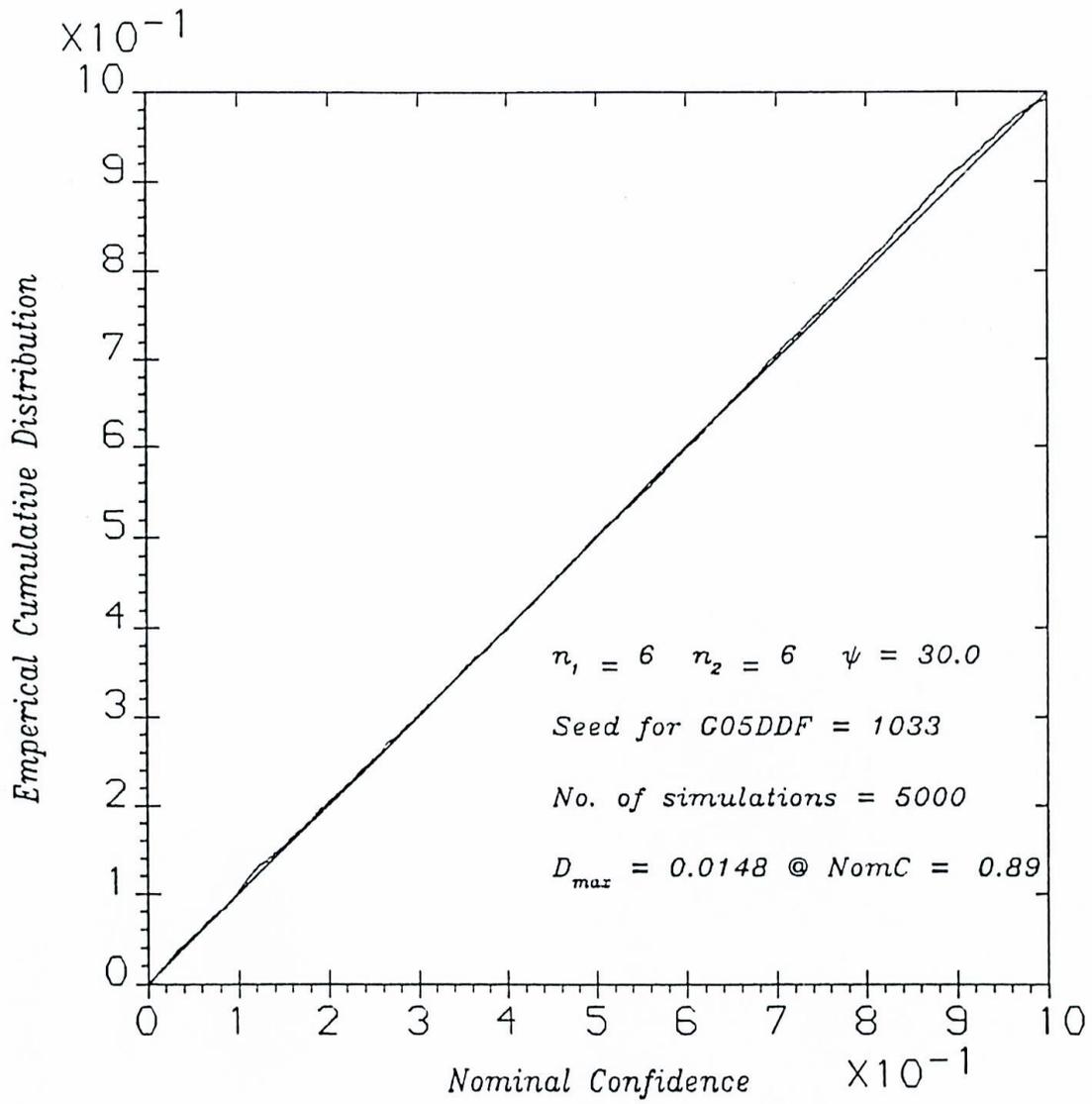

Figure 6.9

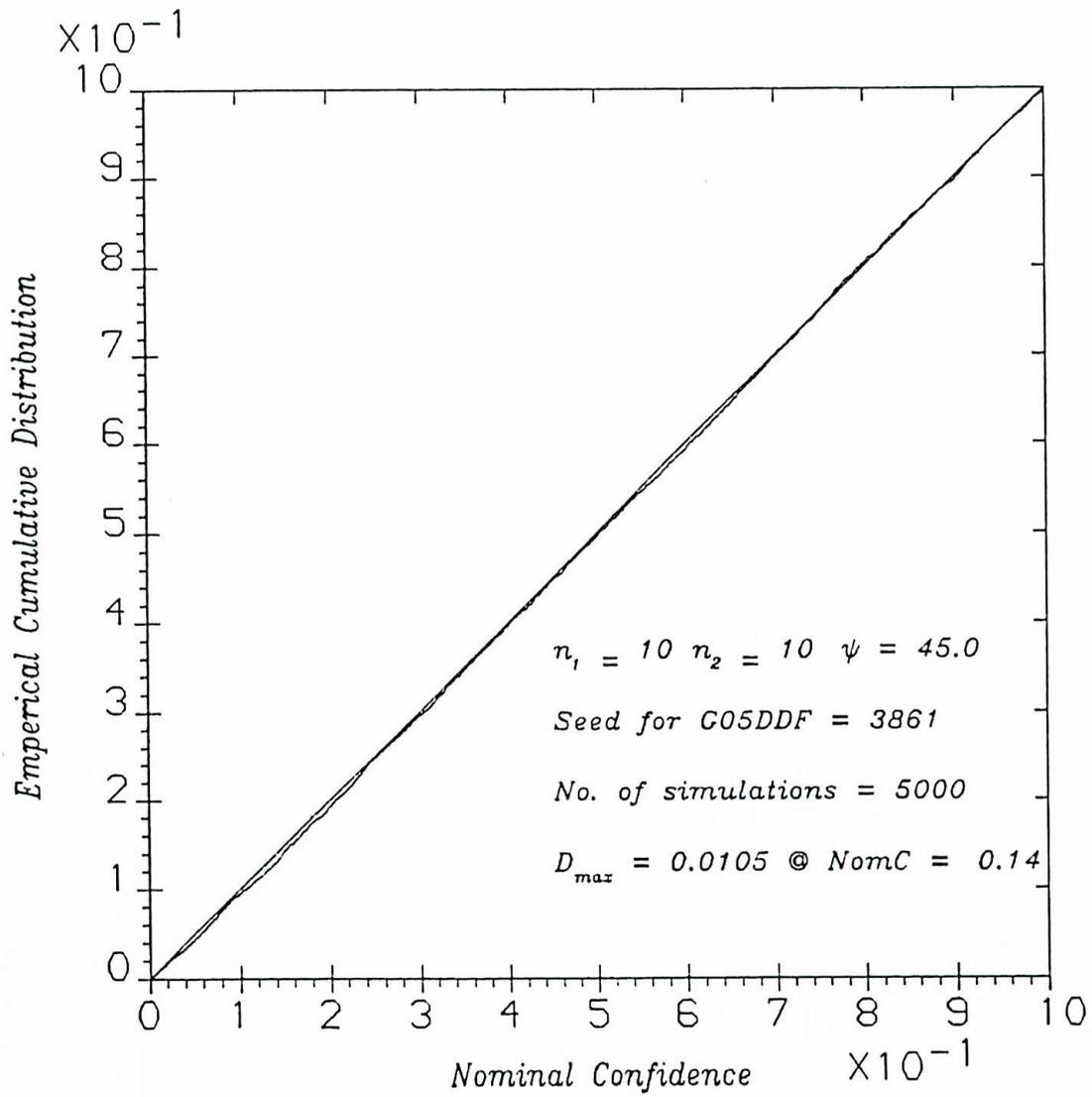

Figure 6.10

A table comparing the powers of $T(1)$ and $V$ using its 'ideal' test criteria, assuming $\zeta$ = 1 in both cases, has been lost (see following postscript), however a synopsis of it has survived. Power was calculated for each $\delta$ = 0.0, 0.5, 1.0, ... , 5.0 . These calculations showed that these statistics have almost identical power functions when $v_1 = v_2$, with Power{$|T(1)|$} - Power{$|V|$} $\leq$ 0.001, whereas the test using $|T(1)|$ is slightly more powerful than the test using $|V|$ when the sample sizes are unequal. The greatest difference between the power functions of the statistics $|T(1)|$ and $|V|$ was when $v_2 = 6$ with $v_1 >> 30$: in this case the power of $|V|$ was found to be always greater than 0.75 times the comparable power of $|T(1)|$. These results could be in part confirmed by use of the 'ideal' criteria in Table 5.2, Appendix 3.

It was intended to calculate the power functions of $T(\zeta)$ and $V$ with its 'ideal' criteria for $\zeta$ = 1/9, 1/4 , 1 , 4 , 9 , but only the case $\zeta$ = 1 was computed; various symmetries were expected in the power functions for these values of $\zeta$

Postscript

All the work presented in this article was established by the present author before 1991. Recently, while undertaking the melancholy task of destroying teaching notes and tutorial solutions of statistics courses once taught at The Queen's University, Belfast, I came across some work I had done on the two-sample problem and its associated computer output. At first I was inclined to destroy these too, but on reflection decided to keep them since publishing on the Internet is easy and allows detailed explanations.  Had I carried out my first instinct the preceding material would have been lost to posterity, if this concept is still valid.

Donald Chambers, 2$^{nd}$ May, 2010.